\documentclass[useAMS,usenatbib]{mn2e}

\usepackage{graphicx}
\usepackage{array}
\usepackage{grffile}
\usepackage{pdfsync}
\usepackage{longtable}
\usepackage{supertabular}
\usepackage{threeparttable}
\usepackage{color}

\newcommand{\lya}[0]{$\mathrm{Ly}\,\alpha$} 
\newcommand{\ion}[2]{{#1}\,{\sevensize {#2}}}
\newcommand{\hion}[2]{{#1}\,{\sevensize {#2}}]}
\newcommand{\fion}[2]{[{#1}\,{\sevensize {#2}}]}
\newcommand{\authornote}[1]{}

\title[Rest-frame far-UV spectroscopy of Himiko] { Deep rest-frame far-UV spectroscopy of the giant Lyman $\alpha$ emitter 'Himiko'} \author[J. Zabl et al.]
	{J.~Zabl$^1$$\thanks{E-mail:johannes@dark-cosmology.dk}$,
	H.U.~N{\o}rgaard-Nielsen$^2$, J.P.U.~Fynbo$^1$, P.~Laursen$^1$,
	M.~Ouchi$^{3,4}$, \newauthor P.~Kj{\ae}rgaard$^5$\\ $^1$Dark Cosmology
	Centre, Niels Bohr Institute, University of Copenhagen, Juliane Maries
	Vej 30, 2100 Copenhagen, Denmark\\ $^2$ National Space Institute (DTU
	Space), Technical University of Denmark, Elektrovej, 2800 Kgs. Lyngby,
	Denmark \\ $^3$ Institute for Cosmic Ray Research, The University of
	Tokyo, 5-1-5 Kashiwanoha, Kashiwa, Chiba 277-8582, Japan \\ $^4$ Kavli
	Institute for the Physics and Mathematics of the Universe (WPI), The
	University of Tokyo, 5-1-5 Kashiwanoha, Kashiwa, \\ Chiba 277-8583,
	Japan \\ $^5$ Niels Bohr Institute, University of Copenhagen, Juliane
Maries Vej 30, 2100 Copenhagen, Denmark } \date{Released XXXX Xxxxx XX}

\pagerange{\pageref{firstpage}--\pageref{lastpage}} \pubyear{2002}

\def\LaTeX{L\kern-.36em\raise.3ex\hbox{a}\kern-.15em
T\kern-.1667em\lower.7ex\hbox{E}\kern-.125emX}

\begin{document}

\label{firstpage}

\maketitle
\begin{abstract}
	
	We present deep 10h VLT/XSHOOTER spectroscopy for an extraordinarily luminous and extended \lya{} emitter at
	$z=6.595$ referred to as \emph{Himiko} and first discussed by
	\citet{Ouchi:2009dp}, with the purpose of constraining the mechanisms
	powering its strong emission. Complementary to the spectrum, we
	discuss NIR imaging data from the CANDELS survey.
	
	We find neither for He\,{\sevensize II} nor any metal line a significant excess, with $3\sigma$ upper limits of $6.8$, $3.1$, and $5.8\times
	10^{-18}\,\mathrm{erg}\,\mathrm{s^{-1}}\,\mathrm{cm}^{-2}$ for \ion{C}{IV}$\,\lambda1549$, \ion{He}{II}$\,\lambda1640$, \ion{C}{III}]$\,\lambda1909$,
	respectively, assuming apertures with  $200\,\mathrm{km}\,\mathrm{s}^{-1}$ widths and offset by $-250\,\mathrm{km}\,\mathrm{s}^{-1}$ w.r.t to the peak \lya{} redshift.

	These limits provide strong evidence that an AGN is not a
	major contribution to \emph{Himiko}'s \lya{} flux. Strong
	conclusions about the presence of \ion{Pop}{III} star-formation or
	gravitational cooling radiation are not possible based on the obtained \ion{He}{II} upper limit.

	Our \lya{} spectrum confirms both spatial extent and flux
	($8.8\pm0.5\times10^{-17}\,\mathrm{erg}\,\mathrm{s^{-1}}\,\mathrm{cm}^{-2}$)
	of previous measurements. In addition, we can unambiguously exclude any
	remaining chance of it being a lower redshift interloper by
	significantly detecting a continuum redwards of \lya{}, while being undetected
	bluewards.

\end{abstract}


\section{Introduction} An increasingly large number of galaxies is found by
their Lyman-$\alpha$ ($\mathrm{Ly}\,\alpha$) emission in narrowband imaging
surveys at redshifts up to $z\sim7.3$
\citep[e.g.][]{Ouchi:2010eo,Shibuya:2012jy}\footnote{Currently, the spectroscopically confirmed LAE with the highest redshift ($z=7.5$, \citealt{Finkelstein:2013uf}) has been found with HST/WFC3 broadband data}.  Searches are ongoing to find
$\mathrm{Ly}\,\alpha$ emitters (LAE) at redshifts $z\sim7.7$ and $z\sim8.8$
\citep[e.g.][]{Clement:2012jm, McCracken:2012gd, 2013A&A...560A..94M},
but first results beyond $z\sim7$ indicate a rapid decline in the fraction of star-forming galaxies with strong observable \lya{} emission \citep[e.g.][]{Konno2014}. This is in agreement with the low number of \lya{} detections in spectroscopic follow-ups for Lyman-break selected galaxies \citep[e.g.][]{Stark2010,Treu2013,Caruana2012,Caruana2014,Pentericci2014}. Such an evolution can be caused either by an increased amount of neutral hydrogen in the vicinity of the galaxies or by a change in galaxy properties, e.g. in the escape of the ionising continuum \citep[e.g.][]{Dijkstra2014b}.
 
Typical LAEs at redshift $z\sim$2\mbox{--}3 are compact and faint
\citep[e.g.][]{Nilsson:2007cx,Grove:2009hw}, but a population of LAEs with emission
extending up to 100 kpc has been found
\citep[e.g.][]{Fynbo:1999bk,Steidel:2000bw,Francis:2001jb,Nilsson:2006hi}.
Currently the most distant object showing characteristics of a $\mathrm{Ly}\,\alpha$ blob (LAB), despite the effects of cosmological surface brightness dimming, is the source \emph{Himiko} found by \citet{Ouchi:2009dp} at a redshift of 6.6 with Subaru/NB921 imaging. 

While low surface brightness extended \lya{} halos are identified to be a
generic property around LAEs \citep[e.g.][]{2011ApJ...736..160S,Matsuda:2012fp}, several mechanisms are theoretically proposed to support the much stronger extended
$\mathrm{Ly}\,\alpha$ emission of LABs.  Each of them might be responsible
either alone or in combination.  The suggested possibilities include cooling
emission from gravitationally inflowing gas \citep[e.g.][]{Haiman:2000di,
Scarlata:2009ce, Dekel:2009fz, Dijkstra:2009ey, FaucherGiguere:2010dd},
superwinds produced by multiple consecutive supernovae
\citep[e.g.][]{Taniguchi:2000ba}, photo-ionisation by AGNs
\citep[e.g.][]{Haiman:2001bb}, extreme starbursts in the largest overdensities,
where the individual galaxies in the proto-cluster are jointly contributing to
make up a blob \citep{Cen:2012ts}, and starbursts within major mergers
\citep{2013ApJ...773..151Y}.

Observational evidence from individual objects suggests that several of these mechanisms may contribute.
 \citet{Hayes:2011bd} find based on polarisation measurements
evidence for $\mathrm{Ly}\,\alpha$ photons to be originating from a central
source and being scattered at the surrounding neutral hydrogen.  In other cases,
evidence for an AGN as a central ionisation source is found directly
\citep[e.g.][]{Kurk:2000ti,Weidinger:2004hb,Bunker2003}.
In a few cases due to the absence of an apparent central ionising source (starburst or AGN) it has been argued for gravitational cooling radiation as the only remaining scenario \citep{Nilsson:2006hi,Smith:2007in}. However, this is a conclusion which can be challenged  \citep[]{Prescott2015a} as halos producing the required amount of cooling radiation would be expected to have a star-forming galaxy at their centre. \citet{Prescott2015a} have found a possible ionising source for the \citet{Nilsson:2006hi} object in a hidden AGN offset from peak \lya{} emission.

Substantial observational efforts have already been devoted to \emph{Himiko}
\citep[e.g.][]{Ouchi:2009dp, 2013ApJ...778..102O, Wagg:2012hu}.  In this paper we
present deep VLT/XSHOOTER \citep{Vernet:2011by} spectroscopy of this remarkable
object, extending over the full range from the optical to the near-infrared (NIR) $H$-band.  The
main purpose of the observation is to search for other emission lines than
\lya{}, which helps to shed further light on the origin of the extended
$\mathrm{Ly}\,\alpha$ emission.  In particular, both a very hot stellar
population \citep[e.g.][]{Schaerer:2002bm,Raiter:2010hs}, as expected for a
metal-free Population III (\ion{Pop}{III}), and gravitational cooling radiation
\citep[e.g.][]{Yang:2006hv} would give rise to relatively strong He\,{\sevensize
II}$\,\lambda1640$ emission.  By contrast, due to preceding metal-enrichment, an
AGN is expected to display in addition to He\,{\sevensize II} high-ionisation
emission lines from C, Si or N.  We supplement the spectroscopic data by analysing CANDELS $J_{F125W}$ and $H_{F160W}$ archival imaging
\citep{Grogin:2011hx,Koekemoer:2011br}.


In section \ref{sec:obs}, we describe our spectroscopic observations,
while we give details about the data reduction in section \ref{sec:datared},
followed by a discussion of the photometry done on archival data (sec.
\ref{sec:photometry}). Results for the \lya{} spatial distribution, the spectral
rest-frame UV continuum, \lya{} flux and profile, and non-detection limits for
the rest-frame far-UV lines are presented in sections \ref{subsec:spatial}
through \ref{sec:detlimits}. 
Subsequently, we discuss in sections \ref{sec:sedfitting} and \ref{sec:furthersed} implications from the broadband
SED, in section \ref{sec:lyaprofile} a possible interpretation of the \lya{} shape,
and finally and most important in section \ref{sec:upperlimitsimpl} the implications from our non-detection limits
for the mechanisms powering \emph{Himiko}.


Throughout the paper a standard cosmology ($\Omega_{\Lambda,0} = 0.7$, $\Omega_{\mathrm{m},0} = 0.3$,
$H_0 = 70\,\mathrm{km}\,\mathrm{s}^{-1}$) was assumed.  All stated magnitudes
are on the AB system \citep{Oke:1974ip}.  Unless otherwise noted, all
wavelengths are converted to vacuum wavelengths and corrected to the
heliocentric standard.  A size of $1\arcsec$ corresponds at $z=6.595$ to a
proper distance of $5.4\,\mathrm{kpc}$.  The universe was at that redshift
$800\,\mathrm{Myr}$ young.  When stating in the following `He\,{\sevensize II}',
`C\,{\sevensize IV}', `C\,{\sevensize III}]', and `N\,{\sevensize V}', we are referring to
He\,{\sevensize II}$\,\lambda1640$,  C\,{\sevensize IV}$\,\lambda\lambda1548,1551$, [C\,{\sevensize III}]C\,{\sevensize III}]$\,\lambda\lambda1907,1909$, and
N\,{\sevensize V}$\,\lambda\lambda1239,1243$, respectively. 


\section{Data}
    
\subsection{Spectroscopic Observations} \label{sec:obs}

\begin{table} \caption{Number of exposures and exposure times per observing block are
		listed. Except in OB1 and OB9, exposures were taken with $1200\,\mathrm{s}$
		both in the VIS and UVB arm.  In the NIR, each of the exposures
		was split into two sub-integrations with half the exposure time
		(e.g. $1200\,\mathrm{s}$ VIS $\Rightarrow$ 2x600$\,\mathrm{s}$ NIR). Where not all
		exposures could be used for the reduction due to passing clouds, both the used and the
		total number are stated. Numbers in square
		brackets indicate the exposure times included in the NIR stack, if different from the VIS stack.}
	\begin{threeparttable} 
    		\begin{tabular}{ccccc}
  
  Date & Obs. Block & \# VIS exp. & exp. time [s] & conditions\tnote{1} \\ 
  \hline \hline
  02.09 &OB\_H\_1-1 & 0/1 & 0/1169.7 & TK \\ 
  &OB\_H\_1-2 & 4 & 4800 & TN/CL \\
  &OB\_H\_1-3 & 2 & 2400 & TN \\
   \hline \multicolumn{2}{r}{Summary 02.09:} & 6/7
  & \multicolumn{2}{l}{ 7200/8369.7 }\\ 
  \hline 03.09 &OB\_H\_1-6 & 4 & 4800 & TN
  \tnote{2}\\ &OB\_H\_1-8 & 4 & 4800 & TN \\
   &OB\_H\_1-9 & 2 & [0]
  1880 \tnote{3} & CL\\
   \hline \multicolumn{2}{r}{Summary 03.09:} & 10/10 &
  \multicolumn{2}{l}{[9600] 11480/11480 }\\
  \hline 04.09 &OB\_H\_1-10 & 6 & 7200
  &  CL \\ &OB\_H\_1-11 & 6 & 7200 & CL \\
   &OB\_H\_1-12 & 2 & 2400 & CL \\
  \hline \multicolumn{2}{r}{Summary 04.09:} & 14 & \multicolumn{2}{l}{16800} \\
  \hline \multicolumn{2}{r}{Complete Summary:} & 30/31 &
  \multicolumn{2}{l}{[33600]35480 ([9.3]9.9hrs)} \\

\hline
\end{tabular} 
\begin{tablenotes}
\item[1] CL: clear, TN: thin cirrus, TK: thick cirrus 
\item[2] TK E $@40^\circ$
\item[3] Two exposures were taken with
	$940\,\mathrm{s}$ (VIS) and 2x$470\,\mathrm{s}$ (NIR). For the
	NIR reduction, we did not use the $470\,\mathrm{s}$ exposures.
\end{tablenotes}
\end{threeparttable}
\label{tab_sci_frames}
\end{table}

\begin{figure} \centering \includegraphics[height =
		4cm]{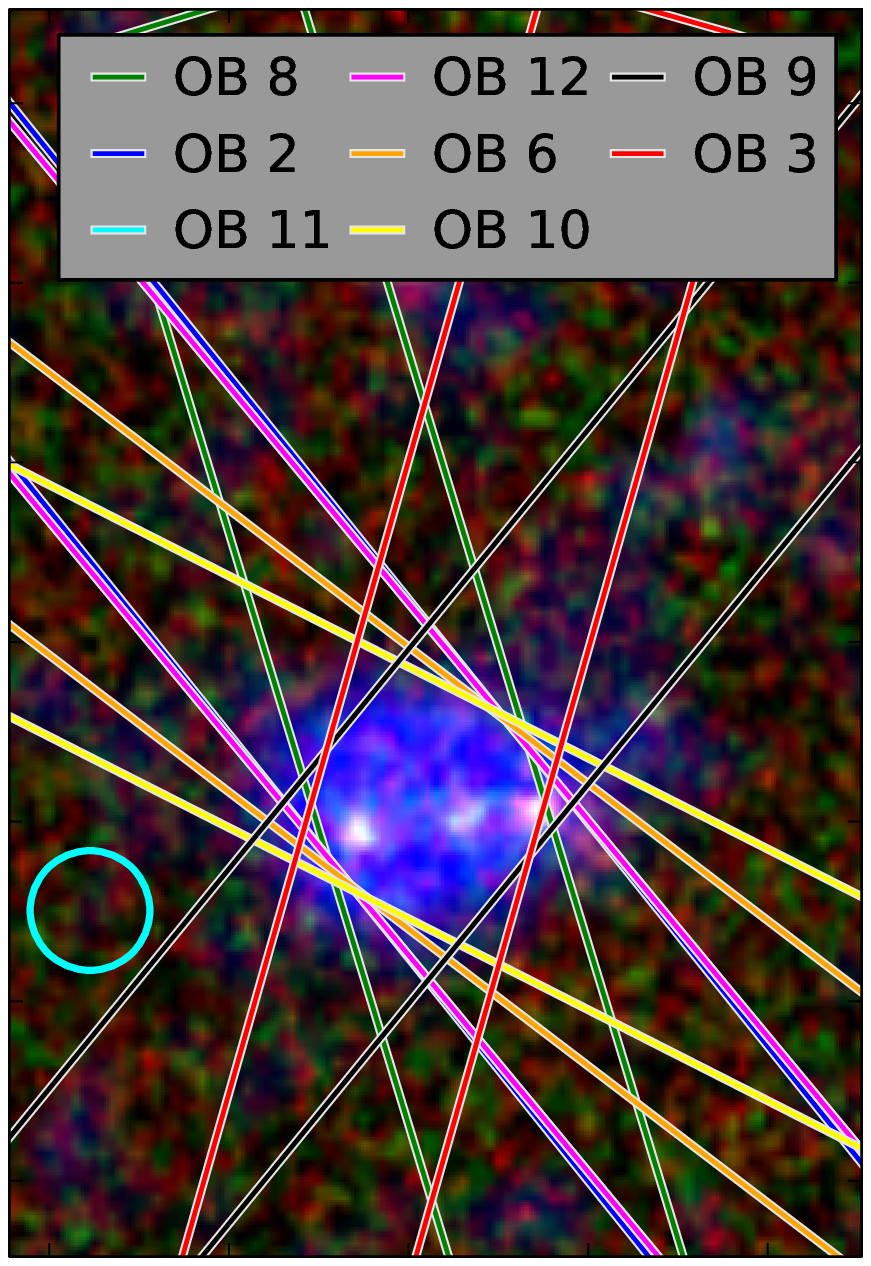}
		\includegraphics[height =4cm]{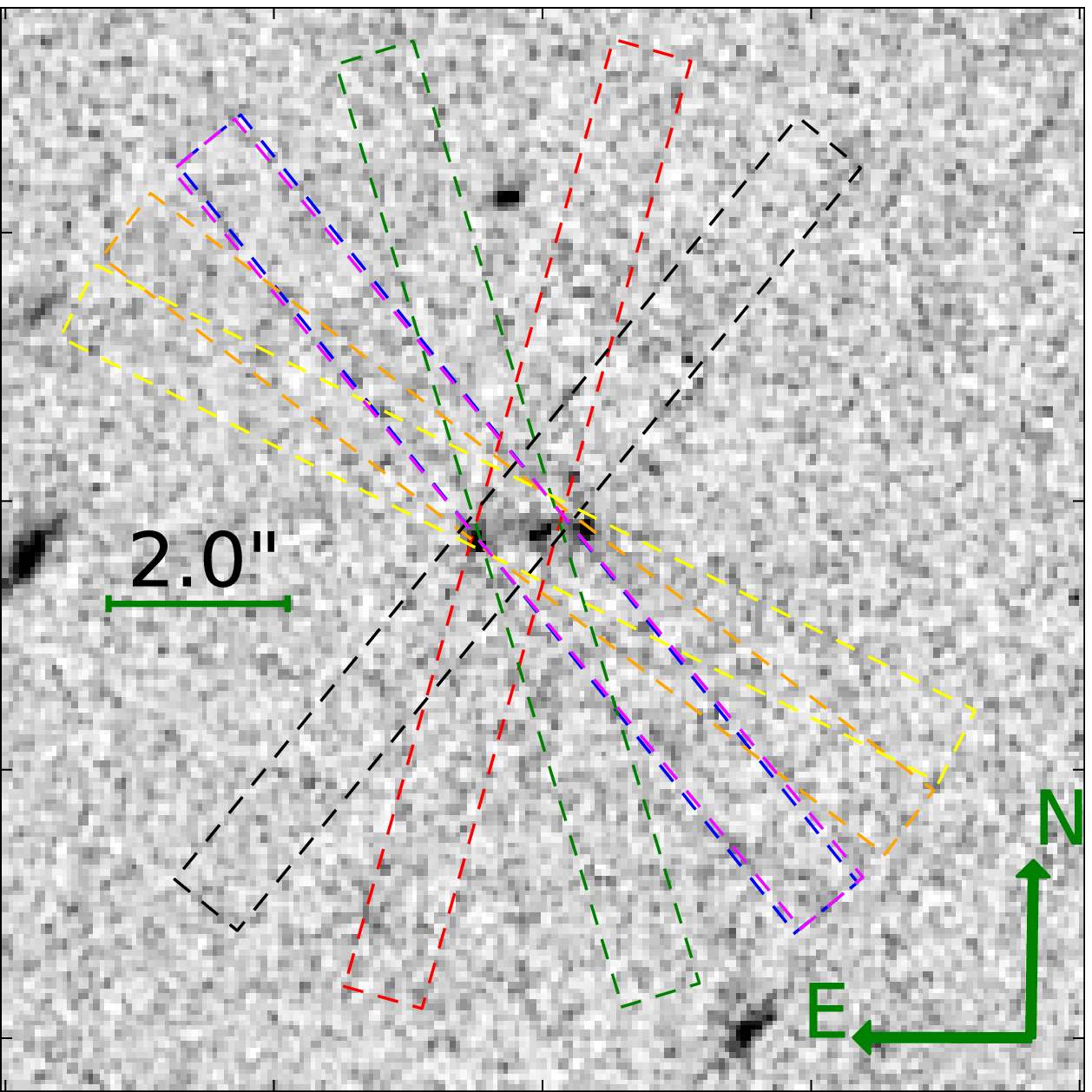} \caption{Alignment
				of the slit in the different OBs w.r.t to
				\emph{Himiko}. A 5\farcs7 x 8\farcs3 colour composite including
				CANDELS WFC3 $H_{F160W}$ and $J_{F125W}$ as red
				and green channels, respectively, and the
				$NB921$ image as blue channel is shown in the
				left panel. $NB921$ is a ground based image with a
				seeing indicated by the
				$0\farcs8$ diameter cyan circle. The 1\farcs5x11\arcsec{} slit, as used in the VIS arm, is included with the different orientations used during the observation. 
				In the right, the 0\farcs9x11\arcsec slits, which were used in the NIR arm,  are overplotted on a 12\farcs1 x 12\farcs1 $J_{F125W}$ cutout.
				The legend lists the OBs in counterclockwise
				order (along columns).
				The positive slit direction is in
				all OBs towards the bottom of the figure, so mainly
				towards the south. 
				OB11 and OB12 cannot be separated
				in the plot, as exactly the same position angles
				were used.  } 
				\label{fig:slit15} 
				\end{figure}

\begin{figure} \includegraphics[width =
		\columnwidth]{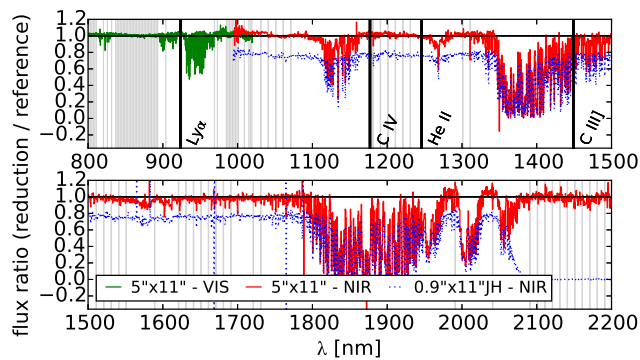}
		\caption{Accuracy of flux calibration based on cross-calibrating
			the spectro-photometric standard LTT7987, taken on September 4, against the
			response function from our main standard Feige110, taken on the night before.  
			The shown curve gives the ratio between the measured LTT7987 flux 
			 and its expectation. In the NIR, results are included both
			 for an observation using the same
			 5\arcsec{}x11\arcsec{} slit for LTT7987 and FEIGE110 and
			 for and observation of LTT7987 with the 0\farcs9x11\arcsec{}JH
			 slit. 
			 Thin vertical lines indicate wavelengths used by the pipeline for fitting a spline to the standard star in the response function calculation.
		 }
	\label{fig:responses} \end{figure}

The XSHOOTER data has been taken at VLT-UT2 (Kueyen) in the second
half of the nights starting on 2011 September 2, 3, and 4, subdivided into nine
different observing blocks (OB) with integration and observing times summarised in Table \ref{tab_sci_frames}. 
We used the same set of slits for XSHOOTER's three spectral arms throughout:
1\farcs{}6x11\arcsec{} (UVB), 1\farcs{}5x11\arcsec{} (VIS), and 0\farcs{}9x11\arcsec{}JH (NIR), which is a slit including
a filter blocking wavelengths longer than $2.1\,\mathrm{\mu m}$, effectively reducing the impact of
scattered light in the NIR spectra. 
All used data was taken under atmospheric conditions classified as either thin cirrus 
(TN) or clear (CL). After excluding OB1 for being affected by thick
clouds, the total usable exposure time was $35480\,\mathrm{s}$.
In our NIR reduction, we used only frames with the same exposure time of $600\,\mathrm{s}$, resulting in a
slightly smaller total time of $33600\,\mathrm{s}$.

Acquisition on our target's narrowband image (NB921; \citealt{Ouchi2008,Ouchi:2009dp}) centroid in the slit's centre was
obtained through a blind offset from a star located
$48\farcs14$ west and $8\farcs99$ north.
We can claim that the pointing accuracy, at least along the slit, was in each of the OBs better
than 0\farcs1, as can be concluded from the spatial centroid of \lya{} in each of
the OBs (cf. Table \ref{tab:lyalpha}).  As the NIR spectrum is observed in XSHOOTER
simultaneously with \lya{} in the VIS arm, we can exclude the possibility of non-detections due to pointing problems.

Spectra were taken with a nod throw of 5\farcs{}0 and a jitter box size of
0\farcs{}5, allowing for an optimal skyline removal. In order to minimise the slit loss, the position angle was set to the parallactic angle at the start of each OB. This is mainly relevant in the NIR,
as the VIS and UVB arms are equipped with atmospheric dispersion correctors.
The corresponding position angles are shown in Fig. \ref{fig:slit15}.

Assuming that emission lines are co-aligned with one or all of the three
continuum bright sources, our decision to use the parallactic angle might have
resulted in a higher than necessary slit loss. This can be seen from Fig.
\ref{fig:slit15} (right). At the time of the observations no HST/WFC3 data was
available.

It is not possible to measure the seeing directly from the
\lya{} spectrum, both due to resonant effects of \lya{} and
 as \emph{Himiko} is not a point source.
Therefore, we needed to extract information about the seeing from the 
FITS header, which has an uncertainty of about 0\farcs2.\footnote{We used the FITS header keyword \emph{HIERARCH ESO TEL IA FWHM}. We tested the use of this keyword with a number of standard and telluric stars. We found that the FWHM values based on this keyword on average agree with a scatter of about 0\farcs2 with the Gaussian FWHM measured from the spatial profile.}
Corrected to both airmass of the observations and the observed wavelength of
$\mathrm{Ly}\,\alpha$ and averaged over sub-integrations, we estimate a seeing in the individual OBs between
0\farcs{}6 and 0\farcs{}8 (FWHM; Table \ref{tab:lyalpha}), with a mean value of
0\farcs{}7 for the stacked OBs, or $0\farcs6$ corrected to the wavelength expected for He\,{\sevensize II}
wavelength.

\subsection{Data reduction} \label{sec:datared}

We performed our final reduction using XSHOOTER pipeline version 2.3.0 \citep{Modigliani:2010hi}, where we made a small modification to the pipeline code. This was to additionally mask all pixels neighbouring those pixels identified by the pipeline as cosmic ray hits. Without this precaution, artefacts remain in the data, which are not indicated in the pipeline quality map.  
Otherwise, we used mainly standard parameters.

The echelle spectra were rectified to a pixel size of
$0.4\,\mathrm{\AA}$ and $1\,\mathrm{\AA}$ in wavelength direction and
0\farcs16 and 0\farcs21 in slit direction for VIS and NIR arm, respectively.
We are only making use of the spectra from XSHOOTER's VIS and NIR arm, as the wavelength range covered by the UVB arm does not contain any information for this object.

While we obtained our final NIR reduction by automatically combining all frames
with the pipeline using the nodding recipe,\footnote{\emph{xsh\_scired\_slit\_nod}} we could improve the result in the VIS reduction somewhat by using our own script.
In the latter case we first reduced the VIS frames in nodding pairs with the
pipeline, and then combined the frames based on a weighted mean, using the
inverse square of the noise maps produced by the pipeline as weights.

\begin{figure} 

\includegraphics[width= 1.1\columnwidth]{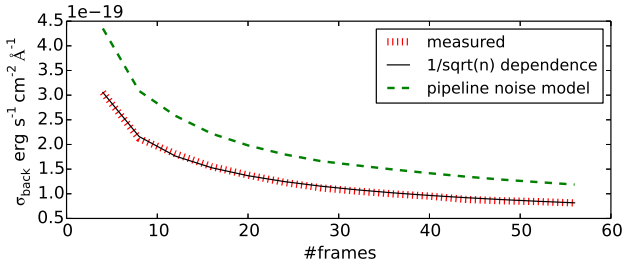} \caption{Pixel
	to pixel noise in skyline free regions close to the expected wavelength
	of He\,{\sevensize II} in the NIR arm. We have successively added 4
	frames corresponding to 2 nodding positions to the stack and calculated
	the $\kappa-\sigma$ ($\kappa = 4$) clipped standard deviation for each
	of the steps. The result is the solid line. The dotted line shows the
	noise level as expected from a $1/\sqrt{N}$ dependence based on the
noise in the first step (4 frames). It is perfectly in agreement with the data.
In addition, the noise prediction from the pipeline noise model is included.}
\label{fig:sndevel} \end{figure}

A nodding reduction is commonly considered as essential for a good skyline
subtraction in the NIR.  Nevertheless, we tried also a stare reduction both in the VIS and the NIR, which could
give in an idealised case a $\sqrt{2}$ lower noise.  
In the case of the NIR spectrum, where the use of dark frames taken with the same exposure time as the science frames is necessary for a stare reduction, we used a large enough number of dark frames to not be limited by their noise.\footnote{As ESO does not take by default enough dark frames for such deep observations, we needed to use frames from several days around the observations for this test. In the nodding reduction used for our result no dark frames are required.}

While the stare reduction worked for the VIS
arm, we experienced in the deep NIR stack spatially abruptly changing residual
structures, which we could not safely remove by modelling with slowly changing functions.
Consequently, we had to decide that a safe stare reduction was not feasible at this point.

By contrast, the pixel to pixel noise decreases as expected with the square root of
the number of exposures in the case of the nodding reduction, as shown in Fig.
\ref{fig:sndevel} for a region around the expected He\,{\sevensize II} line and
using only pixels not affected by skylines and bad pixels. Reassuringly, this indicates that
the structure seen in the stare reduction is at least within the individual
nodding sequences temporarily stable and therefore removed by the nodding
procedure. As the relevant VIS arm wavelength range redwards of \lya{} is affected
by telluric emission lines, which requires a robust sky subtraction, we
decided finally to use a nodding reduction in the VIS arm, too.

In addition to those pixels flagged as bad in the pipeline, we masked skylines,
which we automatically identified by iterative sigma clipping against emission in a stare reduced and non-background
subtracted spectrum, and pixels with unexpected high noise, defined as having absolute
counts larger than 10 times the $1\,\sigma$ error. Except in the region of
\lya{}, where we did not mask these outliers, this assumption is safe in not clipping away any source signal.
Additionally, when determining the S/N within extraction boxes over a certain wavelength region, we excluded those parts of the
spectrum with a noise either 1.5 times or 2.0 times larger than the minimal noise within $200\,\mathrm{\AA}$ of the region's centre.
The decision between 1.5 and 2.0 was made based on the amount of pixels remaining for the analysis. 

Comparing the calculated pixel to pixel noise to the prediction from the
pipeline's noise model, we find for the stack of all 56 NIR
frames a rms noise of
$1.2\times10^{-19}\,\mathrm{erg}\,\mathrm{s}^{-1}\,\mathrm{cm}^{-2}\,\mathrm{\AA}^{-1}$
per pixel, while the direct pixel to pixel variations have a standard
deviation of
$8.1\times10^{-20}\,\mathrm{erg}\,\mathrm{s}^{-1}\,\mathrm{cm}^{-2}\,\mathrm{\AA}^{-1}$.
As we are using the error spectra based on the pipeline throughout
this paper, stated uncertainties for several quantities might be overestimates.
On the other hand, there is correlation in the spectrum due to the rectification, which
is difficult to quantify, especially as it varies with the position in the
spectrum.  A full characterization of the noise would require the propagation of
the covariance matrix \citep[e.g.][]{Horrobin:2008io}, which is currently not
available in the pipeline.

The instrumental resolution at the position of
$\mathrm{Ly}\,\alpha$ and He\,{\sevensize II} was determined based on fitting Gaussians to
nearby skylines.
We get in the two cases R=5300 ($56\,\mathrm{km}\,\mathrm{s}^{-1}$) and  R=5500
($55\,\mathrm{km}\,\mathrm{s}^{-1}$), respectively.

For the determination of the response function, we used a nodding observation of
the standard star Feige110 taken with $5\arcsec$ slits during the night
starting on 2011 September 3, directly before beginning with the observation
of OB\_\emph{Himiko}\_1-6.  We used this response function for all OBs taken within the three nights. 
In the pipeline, the response function is obtained by doing a cubic spline
interpolation through knots at wavelengths having atmospheric transmission close
to 100 per cent.  For ensuring a very good response function close to
\lya{}, we had to remove a knot from the default list at $9270\,\mathrm{\AA}$ and add instead another one at
$9040\,\mathrm{\AA}$.

In order to avoid possible issues of temporal variability of the NIR flat-field illumination, we decided to use the same flat-field observations both for the
standard star and the science frame observations, even though they were taken with
different slits.\footnote{Based on comparing different flat-fields we concluded
that the throughput of the K-blocking filter in the 0.9x11JH slit is in the
J-Band $\ga 97$ per cent.
}
Stability and accuracy of the response function were tested by calibrating
an observation of the flux standard LTT7987, taken in the second night of our program, based on the response
function determined from Feige110 in the first night.  The result of this test is shown in Fig.
\ref{fig:responses}.\footnote{We have used the reference spectra from pipeline version
1.5.0, as those from pipeline 2.3.0 do not allow for cross calibrations of the
same quality.} We can conclude that the accuracy of the spectrophotometry
is about 5 per cent.

In order to reach the maximum possible depth for our science frames, we mainly avoided
spending time on telluric standards. 
Only in the beginning of the observations in the first night we took one telluric standard with the same slit set-up as chosen for our science
observations. We used this frame to fit a model telluric spectrum with ESO's {\sevensize
Molecfit} package \citep{Smette2015}, using the input parameters suggested in
\citet{Kausch2015}. Based on the obtained atmospheric parameters, we created model telluric spectra
for the airmass of each individual nodding pair.
Here, we need to make the strong assumption of
constant atmosphere over a time-scale of several hours, which is unlikely
completely correct. Nevertheless, the obtained accuracy is appropriate for our purpose. 
For the second and third night, we used telluric standards taken for other programs
right before the start of our observations to fit the atmospheric parameters. While these observations were based on
differing slits, we could use the wavelength solution and line kernel obtained
for night one to create appropriate
telluric spectra for nights two and three. 
We adjusted the error spectra after applying telluric corrections.

All 1D spectra were extracted from the rectified 2D frames.  As there is no
detectable trace in the NIR spectrum, we needed to assess the accuracy of the
position of the trace in the rectified frame based on a check on the reduced
standard star.  We find that the centre of the trace does not differ more than
one pixel in slit direction from the
expected position over the complete range of the NIR arm.

\begin{table} \caption{Results of a Gaussian fit to the spatial
		$\mathrm{Ly}\,\alpha$ profiles as measured in the individual OBs
		(cf. Fig. \ref{fig:spatial_lyalpha_profiles}) and the stacked
	spectrum. The 'centre' column gives the displacement w.r.t to the
expected position. In addition, the seeing of the individual OBs, corrected to
observed wavelength and airmass, is stated. } \label{tab:lyalpha}
\begin{tabular}{ccccccccccc} OB & centre  &  FWHM (fit)  &  seeing &
	$f_{\mathrm{Ly}\,\alpha}$ \\ \hline &           & 		  &
	&     {\tiny$10^{-17}$}                       	\\ & '' (slit) &  ''
	(slit)  &  '' & {\tiny$\,\mathrm{erg}\,\mathrm{s}^{-1}\,\mathrm{cm}^{-2}$ } \\ \hline
	
 OB 2 & $-0.01\pm0.04$ & $1.24\pm0.08$ & 0.6 &   $6.1\pm0.5$ \\
 OB 3 & $-0.14\pm0.07$ & $1.43\pm0.16$ & 0.6 &   $6.4\pm0.6$ \\
 OB 6 & $-0.07\pm0.04$ & $1.25\pm0.10$ & 0.8 &   $6.1\pm0.5$ \\
 OB 8 & $-0.03\pm0.05$ & $1.23\pm0.12$ & 0.6 &   $5.6\pm0.4$ \\
 OB 9 & $-0.03\pm0.08$ & $1.51\pm0.20$ & 0.6 &   $6.5\pm0.8$ \\
 OB 10 & $-0.08\pm0.04$ & $1.60\pm0.08$ & 0.8 &   $6.2\pm0.4$ \\
 OB 11 & $-0.05\pm0.02$ & $1.26\pm0.06$ & 0.6 &   $6.2\pm0.4$ \\
 OB 12 & $-0.01\pm0.05$ & $1.20\pm0.11$ & 0.6 &   $6.2\pm0.6$ \\
 all OBs & $-0.02\pm0.02$ & $1.32\pm0.04$ & 0.7 &  $6.1\pm0.2$ \\

\end{tabular} \end{table}

\subsection{Photometry on archival data} \label{sec:photometry}

\begin{figure}

	\begin{tabular}{ccc} 
\includegraphics[width = 0.13\textwidth]{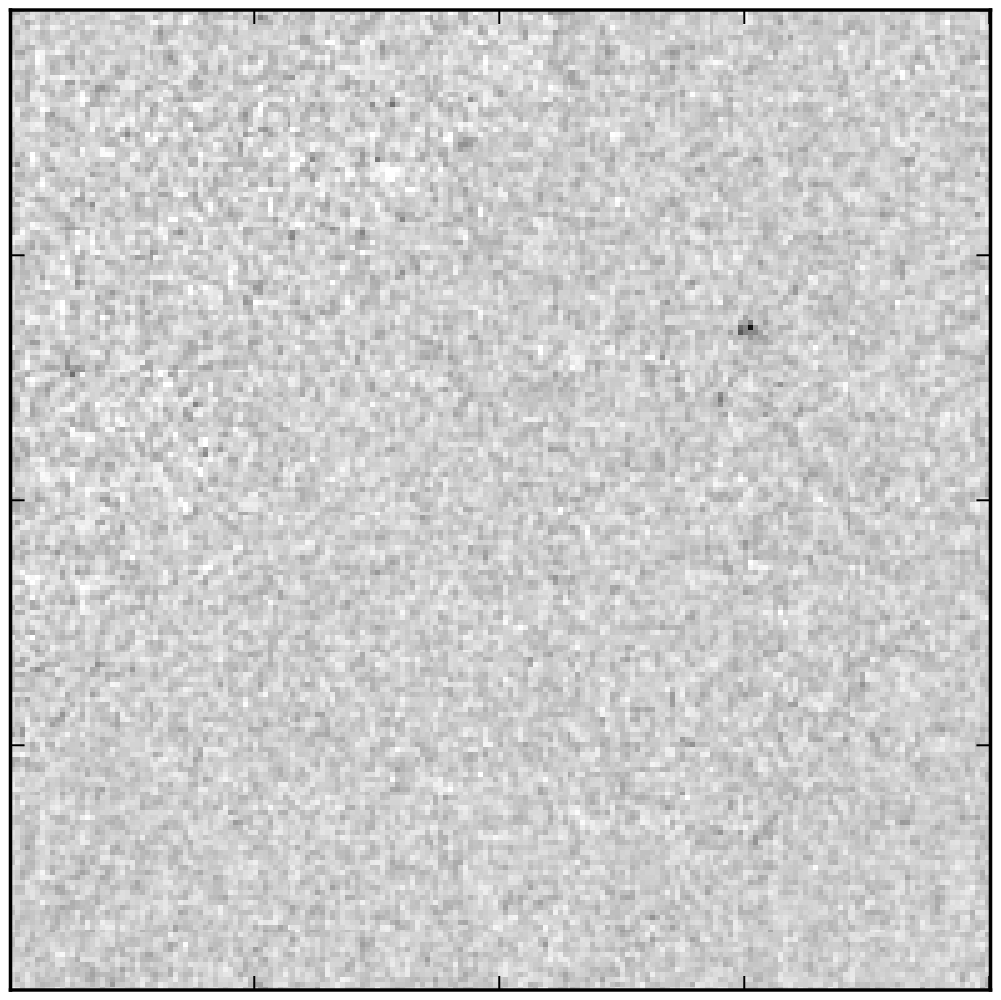} &
\includegraphics[width = 0.13\textwidth]{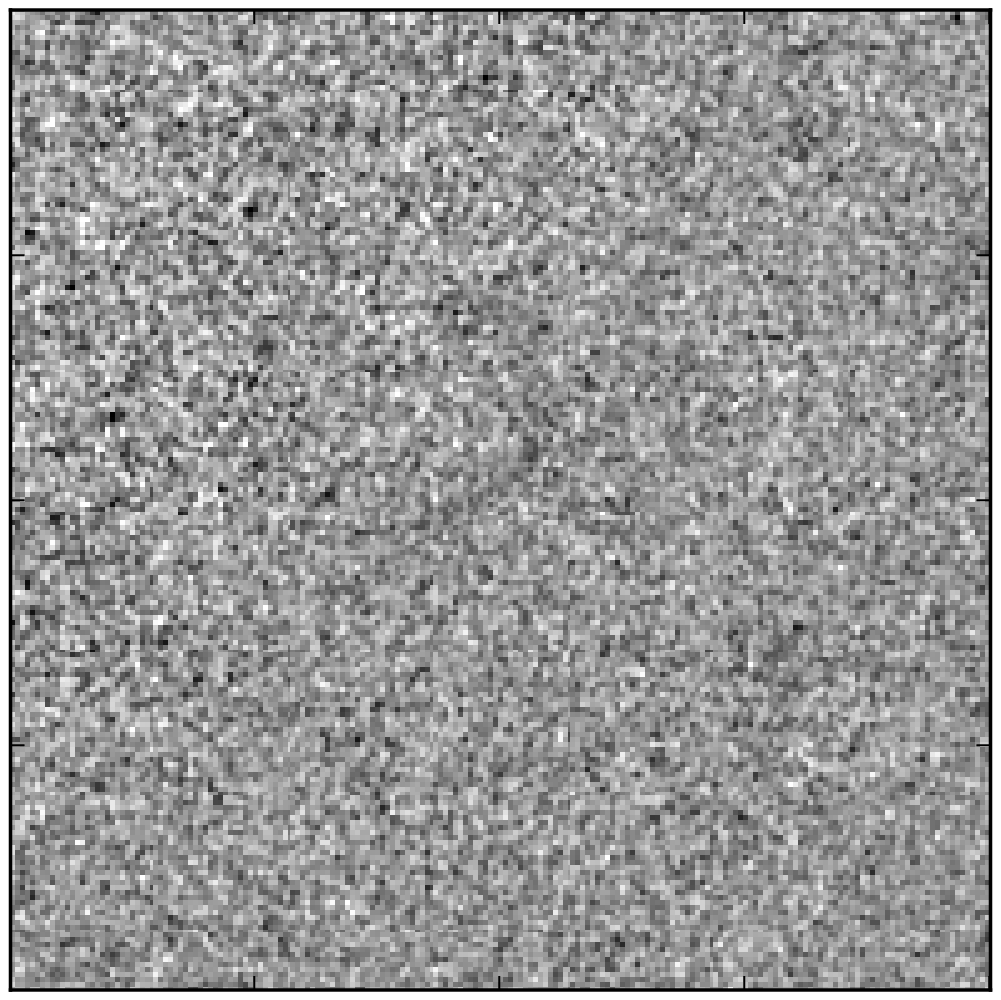} &
\includegraphics[width = 0.13\textwidth]{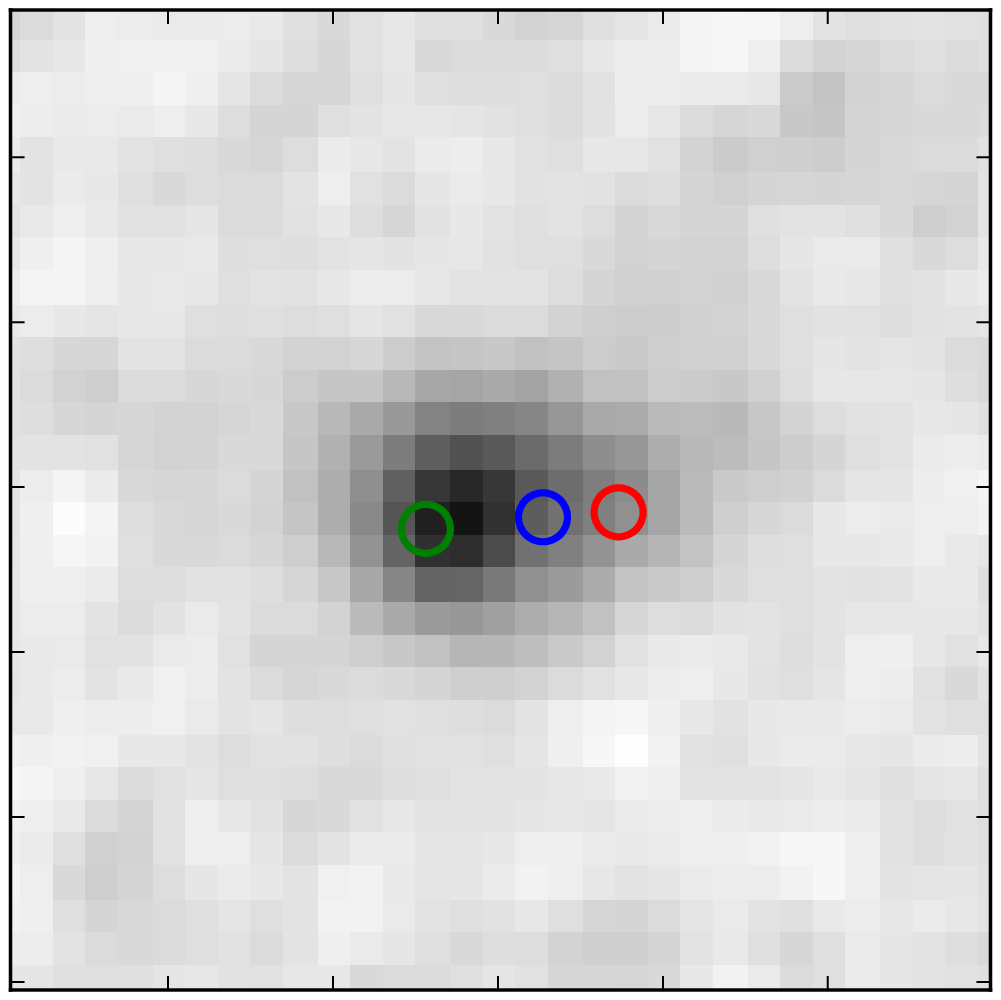} \\
\includegraphics[width = 0.13\textwidth]{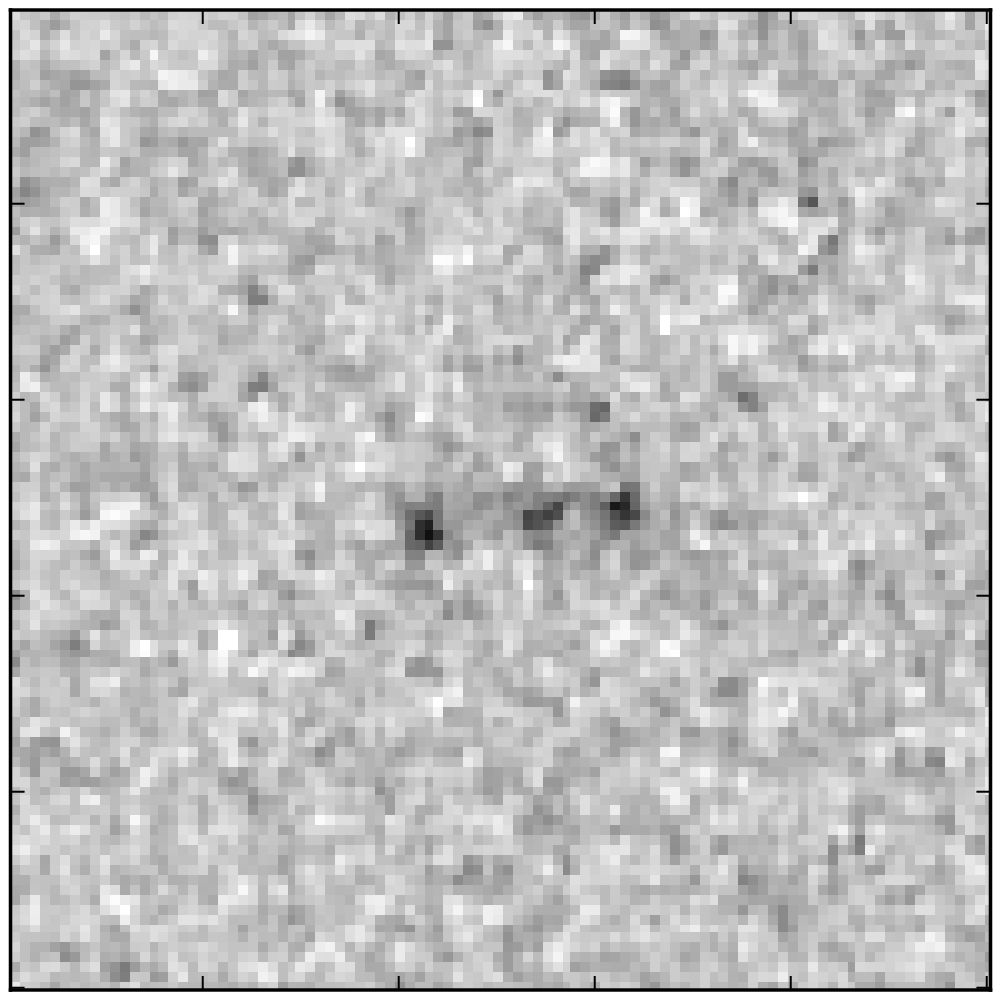} &
\includegraphics[width = 0.13\textwidth]{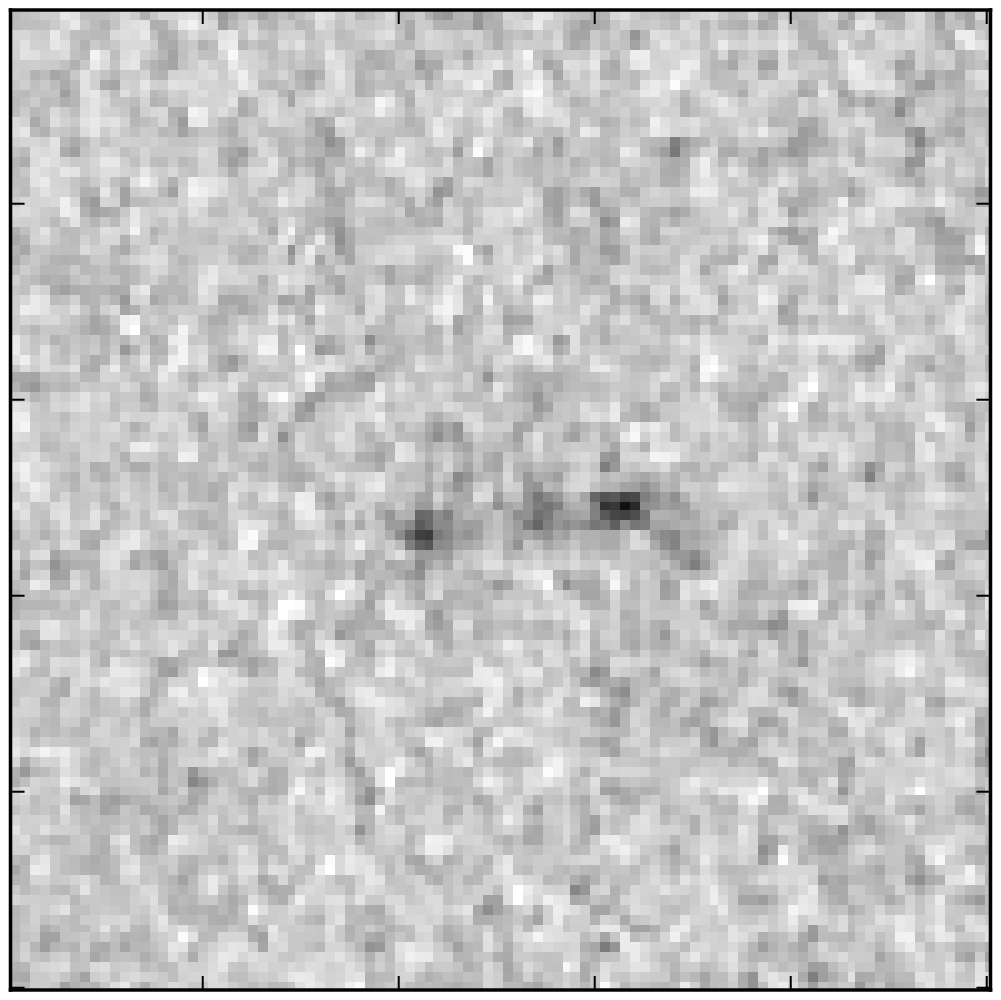} &
\includegraphics[width = 0.13\textwidth]{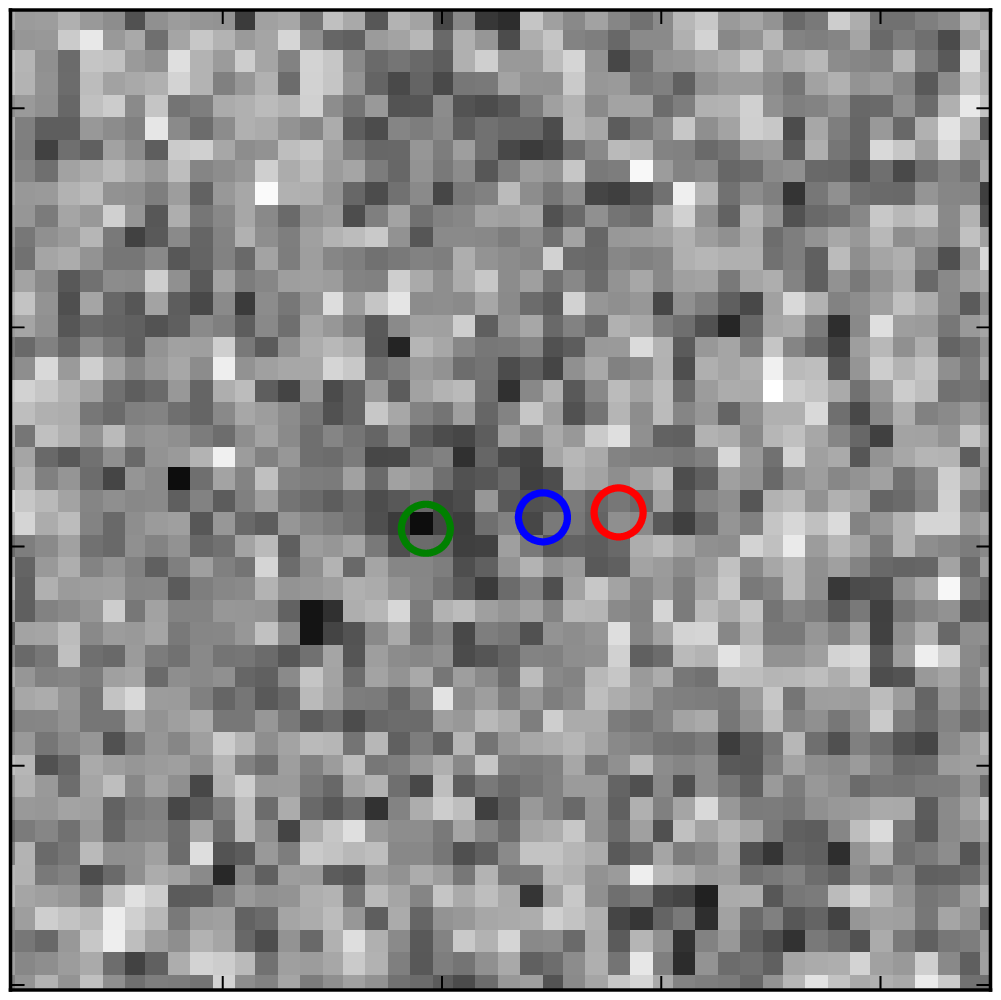} \\

	\end{tabular}

	\caption{ Images in following filters in the stated order: $V_{F606W}$,
	$I_{F814W}$, $NB921$, $J_{F125W}$, $H_{F160W}$,
	UKIDSS/UDS $K$.  The images are scaled
	for optimal viewing from $-3\times \sigma$ to $1.1\times max$, where
	$\sigma$ and $max$ are the standard deviation and maximum value of the
	pixels within a circular annulus around the source, respectively. Green,
	blue, and red circles refer to the three sources visible in the
	$J_{F125W}$ and $H_{F125W}$ images.} \label{fig:cutout} \end{figure} 

As located in the Subaru/\emph{XMM-Newton} deep field, \emph{Himiko} is covered
both by very deep ground and space based imaging, from X-ray to radio, including
the $NB921/Subaru$ narrowband image, which allowed to identify
\emph{Himiko} as a giant $\mathrm{Ly}\,\alpha$ emitter \citep{Ouchi:2009dp}.

Deep \emph{HST} imaging is available from the 
CANDELS survey \citep{Grogin:2011hx,Koekemoer:2011br}.  This includes data from ACS/WFC
in $V_{F606W}$ and $I_{F814W}$ and in $J_{F125W}$ and $H_{F160W}$ with WFC3. We
performed in this work photometry on the CANDELS data. 

It is noteworthy, that recently, both \citet{Jiang:2013vl} and \citet{2013ApJ...778..102O} have published
photometry in $J_{F125W}$ and $H_{F160W}$ based on two other observations
(\emph{HST} GO programs 11149, 12329, 12616 and GO 12265, respectively),
supplemented by deep IRAC 1 \& 2 data from \emph{Spitzer}/IRAC SEDS
\citep{Ashby:2013cc}.  Especially the analysis of \citet{2013ApJ...778..102O} has been
targeted at studying \emph{Himiko}, including additional WFC3/$F098M$
intermediate band data allowing to identify peaks in the \lya{} distribution and ALMA
[C\,{\sevensize II}] observations. 

Ground based near-infrared (NIR) data in \emph{J}, \emph{H}, and \emph{K} is available from the ultra-deep
component (UDS) of the \emph{UKIRT Infrared Deep Sky Survey} (UKIDSS;
\citealt{Lawrence:2007hu}). We are using the UKIDSS data release 8
(UKIDSSDR8PLUS), which has significantly increased depth compared to previous
releases.\footnote{http://www.nottingham.ac.uk/astronomy/UDS/data/dr8.html}

For our SED fitting in section \ref{sec:sedfitting}, we are using NIR photometry determined by us
from the CANDELS and the UKIDSS data and supplement it by the IRAC SEDS (1\&2) and IRAC
SpUDS (3\&4) \citep{Dunlop:2007ue} photometry presented by \cite{2013ApJ...778..102O}.

We determined magnitudes in several apertures on all optical and NIR 
images with {\sevensize SExtractor}'s double image mode \citep{Bertin:1996ww},
with CANDELS
$J_{F125W}$ as detection image.  
Without unwanted resampling, it is due to the
requirement of the same pixel scale in double mode not directly possible to use $J_{F125W}$ as detection image for all other images. Therefore, we put fake-sources in images with the appropriate pixel scale at the positions determined from the $J_{F125W}$ image and used these as input in double mode.
As the UKIDSS photometric system is in VEGA magnitudes, we use the VEGA to AB
magnitude corrections of 0.938, 1.379, 1.900 as stated in \citet{Hewett:2006hy}
for \emph{J}, \emph{H}, and \emph{K}, respectively.  

Consistent with the visual impression (Fig. \ref{fig:slit15}), we detect in the
$J_{F125W}$ image three distinct sources at the position of \emph{Himiko}. We
refer in the following to these sources as $Himiko\mbox{--}E$ (east),
$Himiko\mbox{--}C$ (centre), $Himiko\mbox{--}W$ (west). Their coordinates are
stated in Table \ref{tab:phot_comp}.  The transverse distance between
$Himiko\mbox{--}E$ and $Himiko\mbox{--}W$ is $1\farcs18$
($6.4\,\mathrm{kpc}$), while the distance between $Himiko\mbox{--}C$ and
$Himiko\mbox{--}W$ is $0\farcs46$ ($2.5\,\mathrm{kpc}$).

In order to obtain accurate error estimates for the fluxes measured
within circular apertures,\footnote{SExtractor keywords (FLUX\_APER /
MAG\_APER)} we determined several measurements with apertures of the same size
as used for the object in non-overlapping source free places. 
For images exhibiting correlated noise, as being the case for the drizzled or
resampled mosaic images used for our analysis, this is the appropriate way to account for the correlation.
From the $\kappa$-$\sigma$ clipped standard deviation of these empty-aperture measurements, we obtained the $1\,\sigma$ background limiting flux in the chosen aperture. 
The clipping with a $\kappa = 2.5$ and 30 iterations makes sure that apertures
including strong outlying pixels or which despite the method to find source free
places are not really source free are rejected.

We defined source free regions based on the {\sevensize SExtractor} segmentation
maps and masking of obvious artefacts like spikes or blooming.  In
addition, we made sure that the used regions have approximately the same depth
as the region including \emph{Himiko}.  Depending on the available area, we
had for the different images between 26 and 190 non overlapping empty apertures.

Safely assuming that the noise for the objects is not dominated by the objects, we use
these values as errors on the determined fluxes. All stated
magnitude errors have been converted from the flux errors by the use of:

\begin{equation}
	\sigma_{mag} = 1.0857 \frac{\sigma_{flux}}{flux}
\end{equation}

\begin{table} \caption{ Magnitudes in 0\farcs4 diameter apertures on the
		CANDELS \emph{HST} images are stated for the three distinct
		continuum sources identified with \emph{Himiko} (cf. Fig.
		\ref{fig:slit15}). No aperture corrections are applied.  The
		stated UV slope $\beta$ ($f_\lambda \propto \lambda^\beta$) is
		based on the estimator $\beta = 4.43(J_{F125W} - H_{F160W}) - 2$
		\citep{Dunlop:2012jl}. Upper limits are $1\,\sigma$ values.} \label{tab:phot_comp} \centering
		\begin{tabular}{lrrr} & $Himiko\mbox{--}E$  & $Himiko\mbox{--}C$
			& $Himiko\mbox{--}W$ \\ 
			\hline 
			\hline
			R.A. &  $+2^h17'57''.612$ & $+2^h17'57''.564$  & $+2^h17'57''.533$\\
			Dec & $-5^\circ08'44''.90$  & $-5^\circ 08'44''.83$  & $-5^\circ08'44''.80$ \\ 
			\hline
			$V_{F606W}$ & $>29.53$  & $>29.53$   &  $>29.53$ \\ 
			$I_{F814W}$ & $28.45\pm0.64$ &  $>29.02$  &  $28.52\pm 0.68 $\\
			$J_{F125W}$ &  $26.47\pm 0.07$  &   $26.66\pm 0.10 $ &   $26.53\pm
			0.09$ \\ 
			$H_{F160W}$ &   $26.77\pm 0.12$ & $26.97\pm 0.15$ &   $26.49\pm 0.10$ \\
			\hline

			$J -H$ & $-0.30\pm0.14$  & $-0.31\pm0.18$  & $0.04\pm0.13$ \\
			$\beta$  & $-3.33\pm0.62$  & $-3.37\pm0.80$  &
			$-1.82\pm0.60$ \\

	\end{tabular}

\end{table}

\begin{table}
		\caption{ Magnitudes as measured in $2\arcsec$ apertures for the
		CANDELS/\emph{HST} and the UKIDSS/UDS data. In addition the
		\emph{Spitzer}/IRAC measurements from
		\citet{2013ApJ...778..102O} are included.
		  First and second
		column list measurements without aperture correction and the
		corresponding $1\,\sigma$ errors.  Magnitudes after applying aperture corrections are stated in the
		third column. 
		Upper limits are $1\,\sigma$ values. $\beta$ is calculated as in Table.
		\ref{tab:phot_comp}.  } \label{tab:phot_twoarcsec} 

\centering		
\begin{threeparttable}

		\begin{tabular}{rrrr} Filter & Mag (2'') & $\sigma$ Mag (2'') &
			Total magnitude \\  \hline
			$V_{F606W}$  	&
			$>$27.47	& -- 	&  $>$27.47	 \\
			$I_{F814W}$	&  $>$26.77 	& -- 	&  $>$26.78	 \\ 
			$J_{F125W}$
			&  24.71	& 0.13 	&  24.71         \\ 
			J  	&  25.23 & 0.26	&  25.09 \\ 
			$H_{F160W}$  	&  24.93  & 0.15	&  24.93 \\
			H  	&  $>$25.59 	& -- 	&
			$>$25.44	 \\
			K  	&  24.84 	& 0.22	&  24.72
			\\ 
			3.6$\mu m$ \tnote{1} &  -- 	& 0.09	&  23.69 \\ 
			4.5$\mu m$ \tnote{1}	&  --	& 0.19	&  24.28 \\
			5.8$\mu m$ \tnote{1}	&  -- 	& -- 	&  $>$23.19 \\ 
			8.0$\mu m$ \tnote{1}	&  -- 	& -- &  $>$23.00 \\ 
			\hline $\beta$ & \multicolumn{3}{c}{$-3.0\pm0.9$} \\
			\hline 
		\end{tabular}

	\begin{tablenotes}

	\item[1] Measurements taken from \citet{2013ApJ...778..102O}.
	\end{tablenotes}
\end{threeparttable}
\end{table}

In Table \ref{tab:phot_comp}, measurements within $0\farcs4$ diameter
apertures centred on each of the three peaks are listed for the \emph{HST}
images, while magnitudes within $2\farcs0$ apertures are stated for all used
images in Table \ref{tab:phot_twoarcsec}. In the latter case, the apertures are centred on the
same position as that stated in \citet{Ouchi:2009dp,2013ApJ...778..102O}.  In addition, aperture
corrected magnitudes are included.  To calculate the appropriate aperture corrections for the
images with larger PSFs, we assumed that in all the considered bands the flux is
coming from the three main sources and that the flux ratio between the three peaks
is the same as in the $J_{F125W}$ \emph{HST} image.  This allows us to convolve
this flux distribution with the PSF\footnote{We created fake images with the
	{\sevensize IRAF} task \emph{mkobjects}.} and consequently determine the
	fraction of the total flux included in the aperture on the created fake
	image by using {\sevensize SExtractor} in the same way as for the
	science measurements.  For the PSF profiles, we assumed in the case of
	the UKIRT/WFCAM NIR images Gaussians determined by a fit to nearby
	stars. We get for \emph{J},\emph{H},\emph{K} FWHMs of $0\farcs74$,
	$0\farcs74$, and $0\farcs69$, respectively. 

Finally, Tables \ref{tab:phot_comp} and \ref{tab:phot_twoarcsec} also include
the UV slope $\beta$ ($f_\lambda \propto \lambda^\beta$).  We calculated it
based on the estimator $\beta = 4.43(J_{F125W} - H_{F160W}) - 2$
\citep{Dunlop:2012jl}.  Both the total object and the two eastern components
seem to have very steep slopes of  $-3.0\pm0.9$, $-3.3\pm0.6$, $-3.4\pm0.8$,
respectively. However, the uncertainties are large.

Interestingly, there seems to exist a slight tension between the $J_{F125W}$ in the
data used by us (CANDELS) and that obtained by \citet{2013ApJ...778..102O} with
$24.71\pm0.13$ and $24.99\pm0.08$, respectively.  This corresponds to a
difference of about 1.8\,$\sigma$.  Consequently, they infer a less steep slope
of $\beta = -2.00\pm0.57$.  While \cite{Jiang:2013vl} have not derived
magnitudes for the three individual sources, their total magnitude is with
$24.61\pm0.08$ deviating even more. However, they have been using
{\sevensize SExtractor} \emph{MAG-AUTO} measurements. Therefore, the comparison
between their values and those derived by \citet{2013ApJ...778..102O} and us should be
treated with caution.  On the other hand, the ground based UKIDSS $J$ magnitude is
with $25.09\pm0.26$ closer to the value obtained by \citet{2013ApJ...778..102O}. 

The greatest difference between the CANDELS data and that from
\citet{2013ApJ...778..102O} is in the central component.  They measure from their data
$27.03\pm0.07$ for $J_{F125W}$ and we derive $26.66\pm0.10$.  A subjective visual inspection of the \cite{Jiang:2013vl} data seems to rather confirm the relatively blue colour in the
central blob.

For all stated coordinates, we use the world coordinate system as defined in the
CANDELS $J_{F125W}$ image (J2000).  We find that this coordinate system is
slightly offset w.r.t to the coordinate system used by \citet{Ouchi:2009dp}.
Their coordinate\footnote{RA, DEC = $2^h17'57''.563$, $-05^\circ08'44''.45$ for
	the centroid of the $\mathrm{Ly}\,\alpha$ emission} corresponds to RA,
	DEC = $+2^h17'57''.581$, $-5^\circ08'44''.72$ in the CANDELS $J_{F125W}$  
	astrometric system.  The UKIDSS data appears within the
	uncertainties well matched to the CANDELS astrometry.  Therefore, we do
	not apply any correction here.  Fig. \ref{fig:slit15} (left) includes a
	R,G,B composite using $H_{F160W}$, $J_{F125W}$ (both CANDELS), and $NB921$,
	respectively. Cutouts around the position of \emph{Himiko} for all used
	images are shown in Fig. \ref{fig:cutout}.

\section{Results}


\subsection{Spatial flux distribution and slit losses}
\label{subsec:spatial}

\begin{figure}
	\centering
	\includegraphics[width =
		1.0\columnwidth]{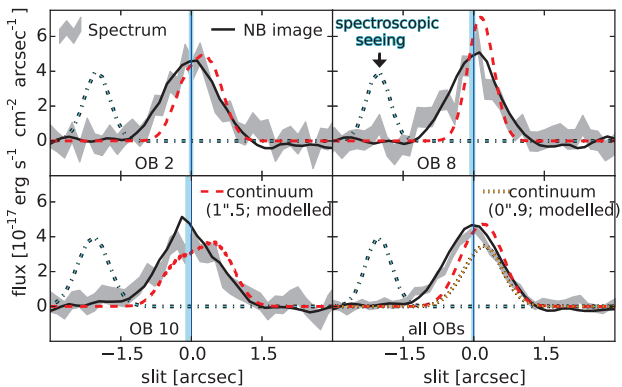}
	\caption{
	\lya{} spatial profiles for three individual example OBs (2, 8 and 10) and for the complete stack of all OBs. Included are profiles directly extracted from the spectrum over the wavelength range from 9227 to 9250 $\mathrm{\AA}$ and profiles extracted from images under the assumption of the 1\farcs{}5 slit. The latter is shown both for the NB921 image and a $J_{F125W}$ image, which was corrected to the seeing of the individual spectroscopic OBs. Finally, for the stack the expected continuum profile is also shown for the 0\farcs9 slit. The legend is split between the different panels, but applies to all four sub-panels. 
		}
		\label{fig:spatial_lyalpha_profiles}
\end{figure}

The different observing blocks (OBs) taken with different position angles allow to compare the spatial extent of \lya{} along the slit to the $NB921$ image for several orientations.
To do so, we extracted for each OB a \lya{} spatial profile averaged over the
wavelength range from 9227 to $9250\,\mathrm{\AA}$, and replicated this for the $NB921$ image by assuming an aperture with the
shape of the $1\farcs5$ slit and calculating a running mean over
pseudo-spatial bins of 0\farcs2.
The 0\farcs8 FWHM seeing of the $NB921$ image
\citep{Ouchi:2009dp} is slightly larger than that expected in the different OBs of our
spectral data (cf. Table. \ref{tab:lyalpha}), but within the uncertainties comparable.

We quantified the centroid and spatial width of \lya{} in each of the individual OBs by fitting Gaussians to the spatial profiles.
Resulting values and the integrated \lya{} flux over the same range are stated in Table \ref{tab:lyalpha}.
The centroids imply an excellent pointing accuracy.

For comparison, the expected spatial profile for the continuum within the slit was estimated based on fake $J_{F125W}$ images (cf. Section \ref{sec:photometry}), convolved with the estimated seeing for each OB.  
All three profiles are shown for three example OBs (2, 8, and 10) in Fig.~\ref{fig:spatial_lyalpha_profiles}.

The profiles extracted from the spectrum are as expected in good agreement with the $NB921$ image.
For those OBs, like OB8, which are nearly perpendicular to the alignment of the three individual sources, the \lya{} profile is significantly more extended than the estimated continuum profile.
The main excess in the emission seems to be towards the north. On the other hand, for those OBs like OB10, with the slit aligned closer to the east-west direction, there seems to be an offset of the continuum towards the west.
This indicates \lya{} emission more concentrated on the eastern parts, in agreement with the F098M/WFC3 observations of \citet{2013ApJ...778..102O}.
We note that the conclusions would not change when assuming different seeing values within $\pm0\farcs2$ of the values taken from the header.

Finally, the spatial profile of the combined \lya{} stack is shown in the lower right panel of Fig. \ref{fig:spatial_lyalpha_profiles}. The shown $NB921$ and continuum profiles have been calculated as the average of all contributing frames with their respective position angles.
In this panel, we are showing in addition the profile for the continuum as expected in the 0\farcs9 slit.
Noteworthy, we expect the continuum offset by 0\farcs2 towards positive slit directions w.r.t to the \lya{} centroid.

Slit losses both for \lya{} and the continuum and both for the 0\farcs{}9 and the 1\farcs{}5 slits were calculated based on the NB921 or the seeing convolved fake $J_{F125W}$ image by determining the flux fraction within slit-like extraction boxes. 
Identical to the profile determination, we treated the individual OBs separately, and simulated the stack, by combining the slit losses for the individual OBs weighted with the appropriate exposure times.

Assuming the 1\farcs5 VIS arm slit and an extraction width of $4\arcsec$ along the slit, being close enough to no loss in slit-direction, results in a slit loss factor of 0.7 (0.4\,mag) for \lya{} based on the NB921 image.

We determined the optimal extraction-mask size for the expected continuum distribution under the assumption of background limited noise. We did so by maximising the ratio between enclosed flux and the square root of the included pixels.
The maximum value is reached in the 1\farcs{}5 VIS slit between 7 and 9 pixels, when keeping the centre of the extraction mask at the formal centre of the slit.
For an eight pixel (1\farcs28) extraction width, we derive a slit loss factor of $0.74\pm0.09$ ($0.32\pm0.13\,\mathrm{mag}$). The stated uncertainties are resulting from the assumed uncertainty on the seeing. 

The spatial distribution for possible He\,{\sevensize II} or metal line emission is not known.
Therefore, we consider hypothetically both the cases that it is co-aligned with the continuum or with the \lya{} emission. Assuming the two distributions, we obtain optimal extraction widths in the 0\farcs9 slit with around six and eight pixels (1\farcs26 and 1\farcs.68), respectively, fixing the trace centre at the formal pointing position in both cases.
As a compromise, we assumed for the default extraction a width of seven pixels, corresponding in the continuum and the $NB921$ case to slit loss factors of $0.53^{+0.10}_{-0.05}$ ($0.7^{+0.11}_{-0.09}$\,mag) and 0.4 (1.0\,mag), respectively.
Certainly, alternative scenarios are possible which could lead to lower or higher slit losses, e.g. if line emission would be originating mainly from the central or the western-most source, respectively.


\subsection{Spectral continuum}
\label{sec:spectral_continuum}

\begin{figure} \centering \includegraphics[width = 0.8\columnwidth
	]{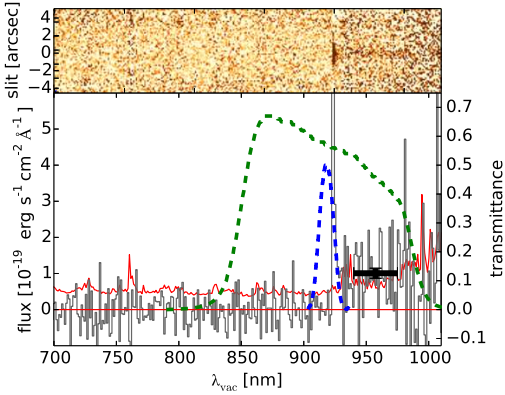} \caption{VIS spectrum telluric corrected and binned to a pixel scale of $11.2\,\mathrm{\AA}\,\mathrm{pixel}^{-1}$.  
	Instead of taking a simple mean of the pixels contributing to the new wavelengths bins, we calculated a weighted mean (sec. \ref{sec:spectral_continuum}). Included are the response profiles for the Subaru
		$NB921$ and $z'$ filters. The trace is shown in black, while the
		red curve is the error on the trace.  
		The black error bar shows the
	continuum level from the spectrophotometry.  No heliocentric velocity correction was applied for this plot.} \label{fig:nb_zprime} \end{figure}

We made the continuum hidden in the noise of the rectified full resolution spectrum visible by strongly binning the telluric corrected 2D spectrum in wavelength direction from an initial pixel scale of $0.4\,\mathrm{\AA}\,\mathrm{pixel}^{-1}$, as produced with the pipeline, to a pixel scale of $11.2\,\mathrm{\AA}\,\mathrm{pixel}^{-1}$.  
Instead of taking a simple mean of the pixels contributing to the new wavelengths bins, we calculated an inverse variance weighted mean, with the variances taken from the pipeline's error-spectrum. This allows to obtain a low resolution spectrum with relatively high S/N in a region strongly affected by telluric emission or absorption. 
One caveat with this approach is that the flux is not correctly conserved for wavelength ranges where both the flux density and the noise changes quickly. This is the case at the blue side of the \lya{} line (cf. Fig. \ref{fig:all_lyalphapos}). Therefore, we used for our binned spectrum a simple mean in the region including \lya{}.

A faint continuum is clearly visible in the resulting spectrum redwards of \lya{} and due to IGM scattering not
bluewards, as expected for \emph{Himiko}'s redshift (Fig. \ref{fig:nb_zprime}).  A similar effort in the
NIR did not reveal an unambiguous continuum detection.

The detection allows to directly determine the continuum flux density close to
\lya{}.  For the interval from 9400 to $9750\,\mathrm{\AA}$ ($1238$ to
$1285\,\mathrm{\AA}$  rest-frame), chosen to be in a region with comparably low noise in our optimally rebinned spectrum,  we get with the $1\farcs28$ extraction mask
a flux density of $10.1\pm1.4\times10^{-20}\mathrm{erg}\,\mathrm{s}^{-1}\,\mathrm{cm}^{-2}\,\mathrm{\AA}^{-1}$
(cf. Fig. \ref{fig:nb_zprime}). 
We used in the calculation a $\kappa\mbox{--}\sigma$ clipping with a $\kappa=2.5$, rejecting one spectral bin.

The determined flux density is equivalent to an observed magnitude of
$25.18\pm0.15$, or after applying the aperture correction of 0.32 mag, of
$24.85\pm0.15$, corresponding to a rest-frame absolute magnitude of $M_{1262;AB} =
-21.99\pm0.15$.
The magnitude is slightly fainter than the the CANDELS $J_{F125W}$ measurement ($24.71\pm0.13$) and slightly brighter than the $H_{F160W}$ magnitude ($24.93\pm0.15$), but within the errors consistent with both of them.

\subsection{$\mathrm{Ly}\,\alpha$} \label{sec:lyalpha}

 
\begin{figure} \center 
		\includegraphics[width =
			0.8\columnwidth]{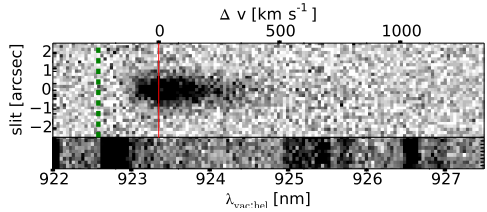}
			\caption{
			2D \lya{} spectrum. The green dashed and solid red vertical lines are at the same
			wavelengths as in Fig. \ref{fig:lyalpha}.  Below, a
		non-background removed spectrum is shown for the same wavelength
	range, indicating positions of skylines.} \label{fig:all_lyalphapos}
\end{figure}

\begin{figure} \centering \includegraphics[width =
		\columnwidth]{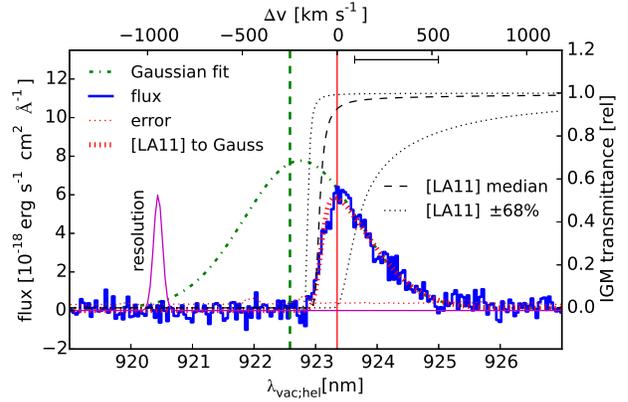} \caption{Extracted
			$\mathrm{Ly}\,\alpha$ spectrum based on the stack of
			all OBs (solid blue). The orange dotted curve gives the
			errors on the flux density. Red solid and dashed green
			vertical lines mark the peak of the
			$\mathrm{Ly}\,\alpha$ line and a velocity offset of $-250\,\mathrm{km}\,\mathrm{s}^{-1}$, respectively. The green
			dot-dashed curve is a fit to the red wing of the
			$\mathrm{Ly}\,\alpha$ line with the centroid fixed as an example to $-160\,\mathrm{km}\,\mathrm{s}^{-1}$. A wavelength
			range as shown by the bar in the upper right corner was
			included in the fit. In addition, both the median and
			the $\pm68\%$ intervals for the IGM transmission from \citet{Laursen:2011ft} [LA11] at this redshift is plotted.
			As a test, we have applied the median transmittance
			to the Gaussian fit, shown as red dashed line.
		Finally, the magenta Gaussian in the left indicates the
	instrumental resolution.} \label{fig:lyalpha} \end{figure}

The final 2D VIS spectrum in the wavelength region around $\mathrm{Ly}\,\alpha$
is shown in Fig. \ref{fig:all_lyalphapos}, from which we extracted the 1D
\lya{} spectrum.  This was done in an optimal way \citep{Horne:1986bg}, using a
Gaussian fit to the spatial profile.
The resulting 1D spectrum is plotted in Fig. \ref{fig:lyalpha}. 
As a sanity check, we compared this result to extractions based on simple apertures of different widths. Large
enough apertures converged within the errors to the result from the optimal
extraction.

We determined several characteristic parameters of the
directly measured $\mathrm{Ly}\,\alpha$ line.  All stated errors are the 68 per cent confidence intervals
around the directly measured value based on 10000 MC random realisations of the
spectrum using the error spectrum.
It needs to be noted that this approach overestimates the uncertainties, as the noise is added twice, once in the actual random process of the observation and once through the simulated perturbations.
This means that the resulting perturbed spectra are effectively representations for a spectrum containing only half the exposure time. Yet, the values based on this simple approach allow to get a sufficient idea of the accuracy of the determined parameters.

The pixel with the maximum flux density is at a $\lambda_{\mathrm{vac;hel}}$ of
$9233.5^{+1.2}_{-0.4}\,\mathrm{\AA}$. This corresponds to a redshift of
$6.5953^{+0.0010}_{-0.0003}$.\footnote{$\mathrm{Ly}\,\alpha$: $\lambda_{rest} =
1215.7\,\mathrm{\AA}$}  The peak $\mathrm{Ly}\,\alpha$ redshift, $z_\mathrm{peak}$, is due to
\lya{} radiative transfer effects likely different from the systemic redshift
of the ionising source (cf. sec. \ref{sec:lyaprofile}).  For the FWHM of the
line we measure $286^{+13}_{-25}\,\mathrm{km}\,\mathrm{s}^{-1}$.
\citet{Ouchi:2009dp} get for their Keck/DEIMOS spectrum a $z_{peak}$ of 6.595
and a FWHM of $251\pm21\,\mathrm{km}\,\mathrm{s}^{-1}$, consistent with our
result. 

Furthermore, we calculated the skewness parameters $S$ and $S_w$
\citep{Kashikawa:2006fi}. For a wavelength range from 9227 to
$9250\,\mathrm{\AA}$ we get values of $0.69\pm0.07$ and
$12.4^{+1.4}_{-1.8}\mathrm{\AA}$ for $S$ and $S_{w}$, respectively.  This is
again consistent with the result obtained by \citet{Ouchi:2009dp}: $S =
0.685\pm0.007$ and $S_w = 13.2\pm0.1\mathrm{\AA}$.  Using an increased
wavelength range from 9220 to $9260\,\mathrm{\AA}$, we measure values of
$0.9\pm0.2$ and $16\pm4\,\mathrm{\AA}$ for the two quantities, being consistent
with the results for the smaller range, but possibly indicating a somewhat
larger value.  Indeed, the spectrum might show some \lya{} emission at high
velocities (cf. Fig. \ref{fig:all_lyalphapos}). However, this is weak enough to
be due to skyline residuals and we do not discuss it further.

Integrating the extracted spectrum over the wavelength range from
$9227\,\mathrm{\AA}$ to $9250\,\mathrm{\AA}$, the obtained flux is
$6.1\pm0.3\times10^{-17}\,\mathrm{erg}\,\mathrm{s^{-1}}\,\mathrm{cm}^{-2}$.
After subtracting the continuum level, the $\mathrm{Ly}\,\alpha$ flux is
$5.9\pm0.3\times10^{-17}\,\mathrm{erg}\,\mathrm{s^{-1}}\,\mathrm{cm}^{-2}$, or
$8.8\pm0.5\times10^{-17}\,\mathrm{erg}\,\mathrm{s}^{-1}\,\mathrm{cm}^{-2}$
after correcting for the slit loss factor of $0.67$, as derived in section
\ref{subsec:spatial}.  This corresponds to a luminosity of
$4.3\pm0.2\times10^{43}\,\mathrm{erg}\,\mathrm{s^{-1}}$.
By comparison, \citet{Ouchi:2009dp} derive a $f_{\mathrm{Ly}\,\alpha}$ of
$7.9\pm0.2\times10^{-17}\,\mathrm{erg}\,\mathrm{s}^{-1}\,\mathrm{cm}^{-2}$ and
$11.2\pm3.6\times10^{-17}\,\mathrm{erg}\,\mathrm{s}^{-1}\,\mathrm{cm}^{-2}$
from the $z'$/$NB921$ photometry and their slit loss corrected Magellan/IMACS
spectrum, respectively.  This is in good agreement with our result, considering
that the slit loss as calculated from the $NB921$ image will only be approximately correct, as the seeing in our observation is not known with certainty (cf. sec.
\ref{subsec:spatial}). 

From the continuum flux measured in the spectrum and the measured \lya{} flux,
we derive a \lya{} (rest-frame) equivalent width ($EW_0$) of
$65\pm9\,\mathrm{\AA}$, nearly identical to the $78^{+8}_{-6}\,\mathrm{\AA}$
stated by \citet{2013ApJ...778..102O}.  \cite{Jiang:2013vl} have in their
recent study derived a \lya{} $EW_0$ of only $22.9\,\mathrm{\AA}$ for
\emph{Himiko}.  The explanation for this discrepancy is that they have fitted a
fixed UV slope based on their relatively blue $J_{F125W}$-$H_{160W}$ and
extrapolated this slope to the position of \lya{}. However, the continuum
magnitude derived from our spectrum is not consistent with this assumption. 


\subsection{Detection limits for rest-frame far-UV lines} \label{sec:detlimits}

\begin{figure*} \centering \includegraphics[width =
		\textwidth]{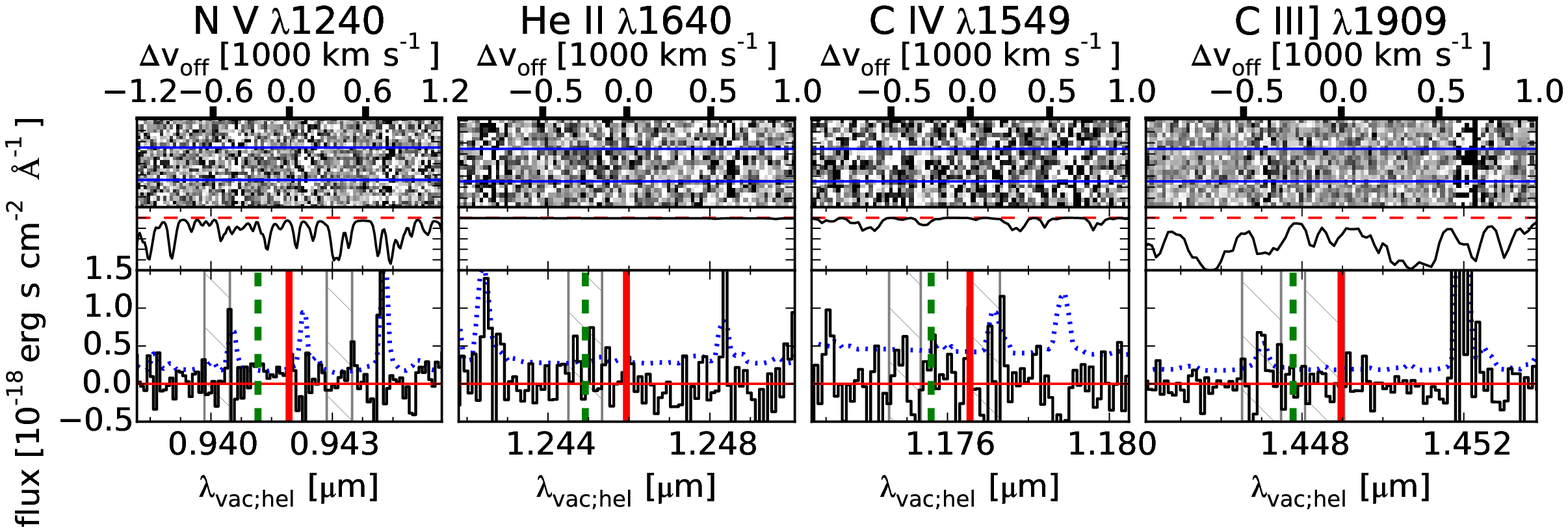}

		\caption{2D and 1D spectra for N\,{\sevensize V}$\,\lambda1240$,
		\ion{He}{II}$\,\lambda1640$, C\,{\sevensize IV}$\,\lambda1549$,
		C\,{\sevensize III]}$\,\lambda1909$. The height of the 2D
		spectrum is 4\arcsec, centred at a slit position of 0\farcs00. The 1D spectrum shows a trace extracted with our default extraction mask. The middle panel shows the telluric 
		absorption as derived from {\sevensize Molecfit} \citep{Kausch2015}, with the red dashed line indicating a transmittance of one and the bottom of the panel being at zero. For the vertical lines
		in the 1D spectrum, compare Fig. \ref{fig:lyalpha}. All
		lines except N\,{\sevensize V}$\,\lambda1240$ are in the NIR
		arm. For the part of the VIS arm spectrum shown for
		N\,{\sevensize V}$\,\lambda1240$, we have  binned the reduced
		spectrum by a factor of two. For the example of our narrower fiducial extraction box ($200\,\mathrm{km}\,\mathrm{s}^{-1}$), the relevant part is marked as hatched region. In the case of the doublets, both relevant regions are indicated. }.
		\label{fig:multline} \end{figure*}

\begin{table} \caption{ $5\,\sigma$ detection limits for N\,{\sevensize
	V}$\,\lambda1240$, C\,{\sevensize IV}$\,\lambda1549$, He\,{\sevensize
	II}$\,\lambda1640$, and C\,{\sevensize III}]$\,\lambda1909$.  They were determined as described in sec. \ref{sec:detlimits}.
	Values are stated in units of
	$10^{-18}\,\mathrm{erg}\,\mathrm{s}^{-1}\,\mathrm{cm}^{-2}$. In addition, we state the flux within the extraction box and the percentage of pixels, which is not excluded in
	our bad-pixel mapping. Continuum and telluric absorption is corrected.
     The three different redshift/width combinations refer to peak
	and FWHM of the measured $\mathrm{Ly}\,\alpha$ line, and two fiducial masks
	used as examples. The symbols refer to those
	included in Fig. \ref{fig:heii_detlimit}.} \label{tab:detlimit}
	\centering
    \begin{threeparttable}

	\begin{tabular}{cccc}  
		
		$\Delta v$ \tnote{1}	& 0 & -250 &  -250 \\
		$width$ \tnote{1} & 286 &  600 & 200 \\
		Symbol & cross & circle & diamond \\
		\hline
		\hline
		
		N\,{\sevensize V}   & 	6.9  & 	9.2  & 	4.5  \\
				    & 	1.1 [40\%] & 	0.6 [46\%] & 	0.2 [52\%] \\
		He\,{\sevensize II} & 	5.7  & 	7.9 & 	5.1  \\
				    &   0.2 [100\%] & 	1.9 [100\%] & 	1.3 [100\%] \\
		C\,{\sevensize IV}  & 	14.3  & 	19.7  & 	11.4 \\
 				    & 	1.0 [69\%] & 	0.1 [65\%] & 	-0.3 [68\%] \\
		C\,{\sevensize III}] &  9.5  & 	14.1  & 	9.7 \\
				     &	-1.5 [82\%] & 	-0.7 [59\%] & 	-0.3 [60\%] \\

	\end{tabular}
    
    	\begin{tablenotes}
	\item[1] [$\mathrm{km}\,\mathrm{s}^{-1}$]
	\end{tablenotes}

\end{threeparttable}

\end{table}

\begin{figure}
	
	\includegraphics[width = \columnwidth]{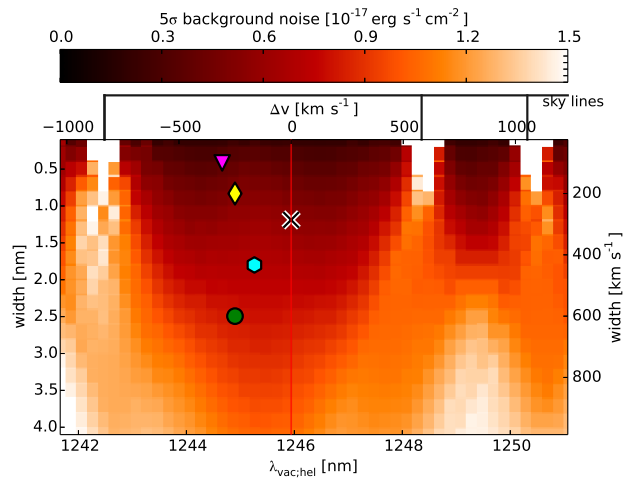}
	\caption{ $5\sigma$ detection limits in a region around He\,{\sevensize
	II} in extraction boxes of different widths and placed at different
	wavelengths corresponding to different velocity offsets w.r.t to the
	peak \lya{} redshift of 6.595, which is marked by a red vertical line.  
	For all boxes, the extent of the box in slit direction was 7 pixel.
	Further details are given in sec. \ref{sec:detlimits}.
	The cross marks redshift and width of the \lya{} line, while the yellow diamond and the green circle indicate our two fiducial boxes. Finally, the magenta triangle and the cyan hexagon indicate the two boxes giving the highest positive \ion{He}{II} excess. 
 } \label{fig:heii_detlimit}
	
\end{figure}

As we do not unambiguously detect any of the potentially expected emission lines
except \lya{}, we focus on determining accurate detection limits.  The
instrumental resolution of $55\,\mathrm{km}\,\mathrm{s}^{-1}$ allows to resolve
the considered lines for expected line widths. 
Therefore, detection limits depend strongly both on the line width, and, due to the large number of skylines of different
strengths, on the exact systemic redshift.  As the
systemic redshift is not known exactly from the \lya{} profile alone and the
search for \fion{C}{II} by \cite{2013ApJ...778..102O} resulted in a non-detection, too, all limits
need to be determined over a reasonable wavelength range.

Studies to determine velocity offsets between the systemic redshift and 
\lya{} for LAEs around $z\sim2\mbox{--}3$ find \lya{} offsets towards the red ranging between
$100\,\mathrm{km}\,\mathrm{s}^{-1}$ and $350\,\mathrm{km}\,\mathrm{s}^{-1}$
\citep[e.g.][]{McLinden:2011ek,Hashimoto:2013cs,2013A&A...551A..93G,Chonis:2013hi},
while the typical velocity offset for $z\sim3$ LBGs is with about $450\,\mathrm{km}\,\mathrm{s}^{-1}$ higher \citep{2010ApJ...717..289S}. 
On the other hand, in LABs also small negative offsets (blueshifted \lya{})
have been observed \citep{McLinden:2013fw}.

Whereas offsets at the redshifts of the aforementioned studies are produced by
dynamics and properties of the interstellar (ISM) and circumgalactic medium
(CGM), at the redshift of \emph{Himiko} an apparent offset can
also be produced by a partially neutral IGM (cf. sec. \ref{sec:lyaprofile}).

Motivated by the lower redshift studies, we formally searched for emission from
the relevant rest-UV lines for $-1000\,\mathrm{km}\,\mathrm{s}^{-1}<\Delta v <
1000\,\mathrm{km}\,\mathrm{s}^{-1}$ from peak \lya{}.  We
calculated for this range the noise, $\sigma_\mathrm{m}$, for extraction boxes with widths up to  
$1000\,\mathrm{km}\,\mathrm{s}^{-1}$. The boxes' height in spatial direction
was in all cases corresponding to the 1\farcs47 trace.

 \begin{equation} \sigma_{m} = \sqrt{\sum_{\mathrm{i},\mathrm{j} \notin
	 \textnormal{bp-mask}}\sigma^2_{\mathrm{i},\mathrm{j}}}
	 \underbrace{\times
		 \;\;(\;1/f_{\mathrm{included}})}_{\textnormal{applied when
		 stating detection limits}} \label{eq_error_calc} \end{equation}
		 Here, $\sigma_{i,j}$ are the noise values for the individual
		 pixels from the pipeline's error model.
Skylines, and bad and high-noise pixels, determined as described in sec.
\ref{sec:datared}, were excluded in the sum, effectively leaving a certain
fraction $f_\mathrm{included}$ of a box's pixels. 
The signal-to-noise (S/N) was obtained by dividing the spectrum's flux
integrated over the same non-excluded pixels by the determined noise. When
stating detection limits, we rescaled the calculated noise by the inverse of the
fraction of included pixels (cf. eq.  \ref{eq_error_calc}). 

In the case of the N\,{\sevensize V}$\,\lambda\lambda 1238,1242$, C\,{\sevensize IV}$\,\lambda\lambda 1548, 1551$, and [C\,{\sevensize III}],\,C{\sevensize III}]$\,\lambda\lambda 1907, 1909$ doublets we formally calculated values jointly
in two boxes centred on the wavelengths of the two components for a given
redshift.  Wide extraction boxes merge into a single box.  Where we state
wavelengths instead of redshift or velocity offset, we refer to the
central-wavelength of the expected 'blend'. Widths are stated for the individual boxes, meaning that the effective box widths are larger.

While we test over a wide parameter space, we refer in the following several
times to somewhat arbitrary fiducial detection limits based on a narrow $200\,\mathrm{km}\,\mathrm{s}^{-1}$
and a wider $600\,\mathrm{km}\,\mathrm{s}^{-1}$ extraction box, assuming a systemic redshift $250\,\mathrm{km}\,\mathrm{s}^{-1}$
bluewards of peak \lya{}, which is as mentioned above a typical value for LAEs. This redshift is also marked in
several plots throughout the paper as green dashed line.

We took account for the continuum by removing an estimated continuum directly in the telluric corrected rectified 2D frame.
The assumed spatial profiles in cross dispersion direction for the $0\farcs9$ NIR and
$1\farcs5$ VIS slits are those estimated from the seeing-convolved \emph{HST}
imaging (cf. sec. \ref{subsec:spatial}).
While we assumed for the region around N\,{\sevensize V}$\,\lambda1240$ in the VIS arm a continuum with
$f_\lambda(\lambda) =
10.2\times10^{-20}\,\mathrm{erg}\,\mathrm{s}^{-1}\,\mathrm{cm}^{-2}\,\mathrm{\AA}^{-1}$,
being the flux measured directly from the spectrum as described in
sec.~\ref{sec:lyalpha} and corrected for aperture loss,  
we were using for the NIR spectrum a $f_\lambda(\lambda)=3\times
10^{-20}\,\mathrm{erg}\,\mathrm{s}^{-1}\,\mathrm{cm}^{-2}\,\mathrm{\AA}^{-1}$
within the slit at the effective wavelength of $H_{160W}$ and a spectral slope of $\beta = -2$.

The used NIR flux density is close to that of the measured $H_{F160W}$.
Due do the difference between our $J_{F125W}$ and the measurement based on UKIDSS $J$ and the $J_{F125W}$ by \citet{2013ApJ...778..102O}, we decided for the conservative option\footnote{As the $J_{F125W}$ magnitude of \citet{2013ApJ...778..102O} is fainter, this results in higher upper limits both for the line fluxes and the $EW_0$.} not to follow the profile shape seen by our data and use a continuum flat in $f_\nu$ instead, even so the corresponding $\beta=-2$ is only at the upper end of the uncertainty
range allowed from the measurement in our work (cf. also sec. \ref{sec:photometry}).

In Table \ref{tab:detlimit}, $5\,\sigma$ detection limits, extracted fluxes, and
the fraction of non-rejected pixels are stated all for N\,{\sevensize V},
C\,{\sevensize IV}, He\,{\sevensize II}, and C\,{\sevensize III}] in three
different extraction apertures.  The spectra for the relevant regions are
shown in Fig. \ref{fig:multline}.


\subsection{He\,{\sevensize II}} \label{sec:heii} 
\begin{figure}
	\includegraphics[width =
		\columnwidth]{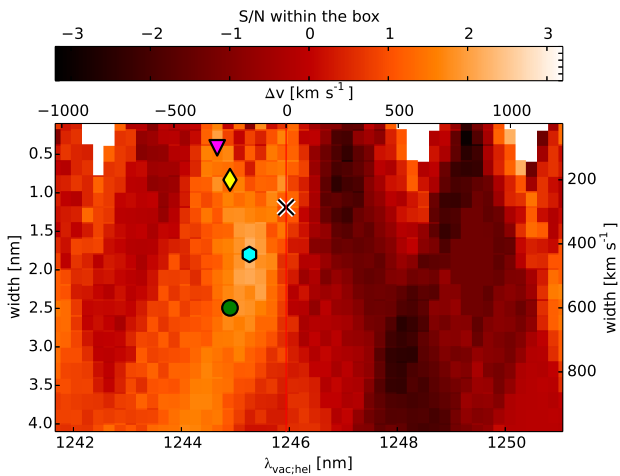} \caption{
		Signal-to-Noise ratio (S/N) measured within extraction boxes
		equivalent to those included in Fig. \ref{fig:heii_detlimit}.
		No continuum and telluric correction was applied for this plot. 
		Maximum values are reached for boxes with $\lambda_{hel;vac} =
		12453\,\mathrm{\AA}$ with a velocity width of
		$430\,\mathrm{km}\,\mathrm{s}^{-1}$ (cyan hexagon) and
		$\lambda_{hel;vac} = 12447\,\mathrm{\AA}$ with a velocity width
		of $100\,\mathrm{km}\,\mathrm{s}^{-1}$ (magenta triangle),
		respectively.}
		
	\label{fig:heii_sn}
\end{figure}

\begin{table}
	\centering
	\caption{2D spectrum and fake-source analysis for a wavelength range corresponding to the expected \ion{He}{II} wavelength. In the row \emph{observed}, the 2D spectrum in the region around the possible He\,{\sevensize II} excess is shown. 
    The grayscale extends linearly between $\pm1\times 10^{-19}
				\,\mathrm{erg}\,\mathrm{s}^{-1}\,\mathrm{cm}^{-2}\,\mathrm{\AA}^{-1}$
				(white $<$ black). 
				Above a Gaussian smoothed version is shown with the same scale, at the top of which a sky-spectrum for the same region is included. Both the \emph{observed} and \emph{smoothed} images are identical in both columns.  The magenta dashed horizontal lines indicate our default extraction mask. 
				 Below the \emph{observed} row, images are shown, where Gaussian fake lines have been added with different strengths for two different FWHM and wavelength combinations. Widths and positions are indicated by the cyan vertical lines. The integrated flux in the respective fake lines is stated in the leftmost column.}
	\begin{threeparttable}
	\begin{tabular}{cc@{\hskip 0.5Ex}c}
	\hline
	   flux\tnote{1} & 100 $\mathrm{km}\,\mathrm{s}^{-1}$ & 400 $\mathrm{km}\,\mathrm{s}^{-1}$ \\
	    \hline
	    \hline	
smoothed	& 	\includegraphics[width=0.36\columnwidth]{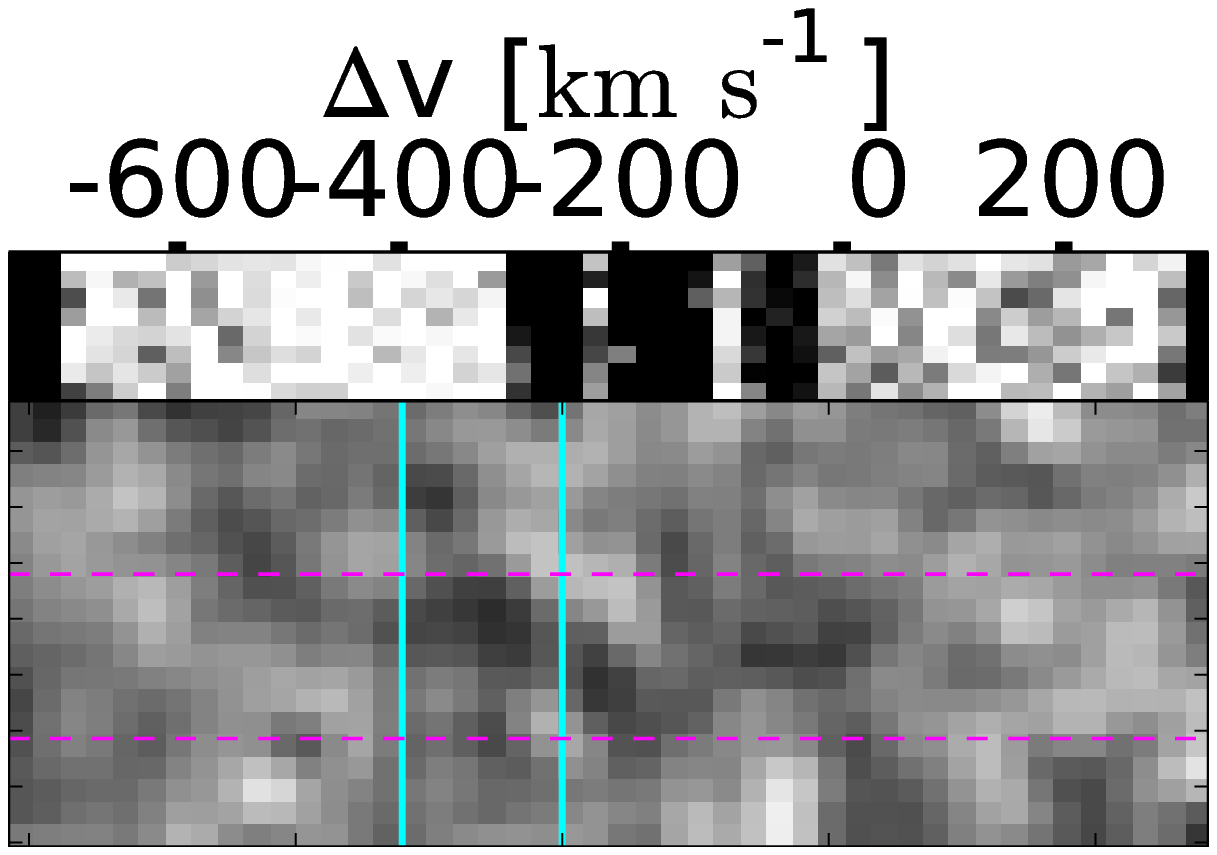} &
		\includegraphics[width=0.36\columnwidth]{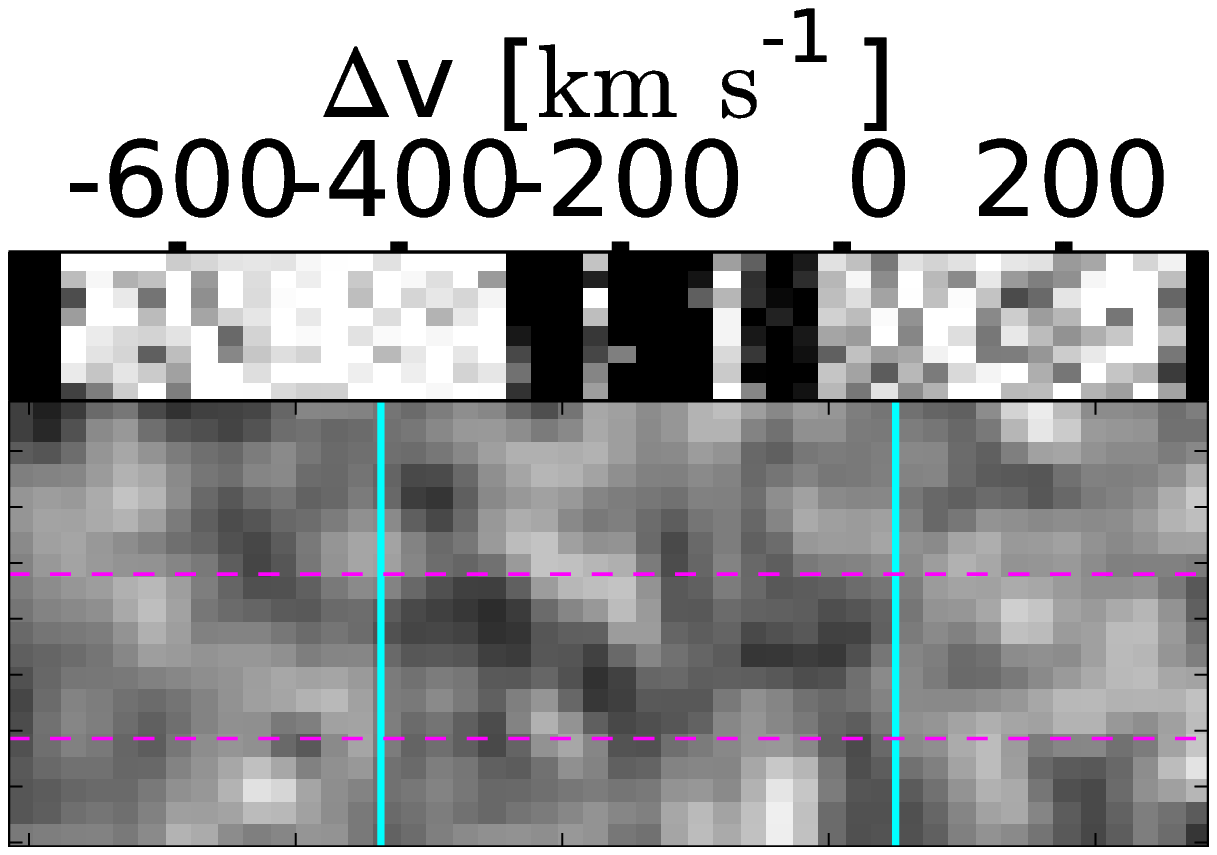}
				\\ 
				\hline
observed	& 	\includegraphics[width=0.36\columnwidth]{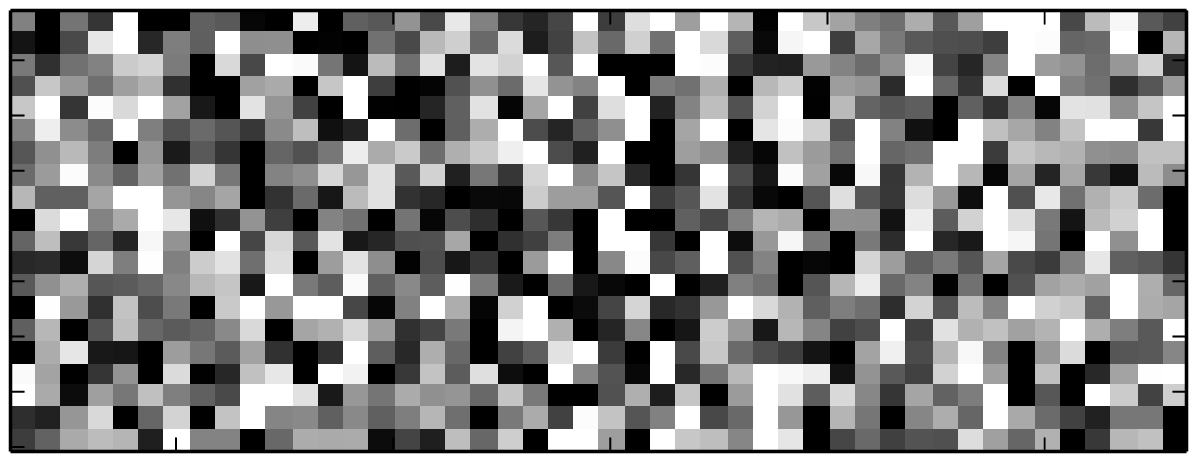} &
		\includegraphics[width=0.36\columnwidth]{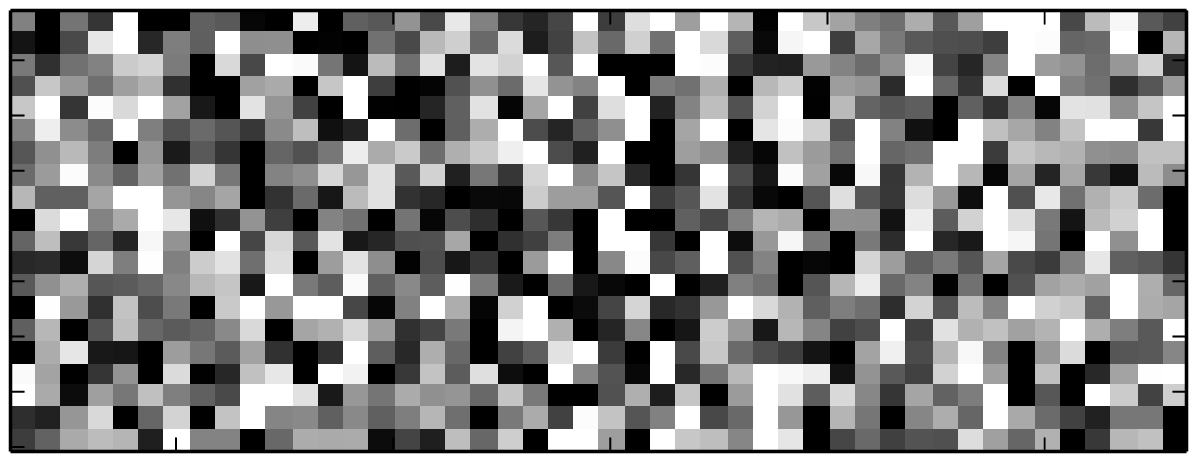}
				\\ 
				\hline
+2.0 & \includegraphics[width=0.36\columnwidth]{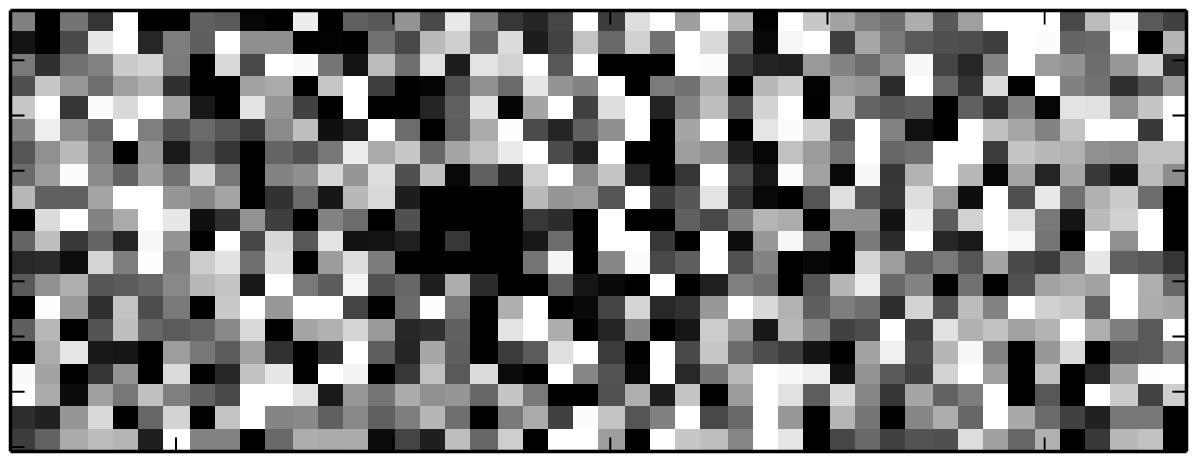} &
		\includegraphics[width=0.36\columnwidth]{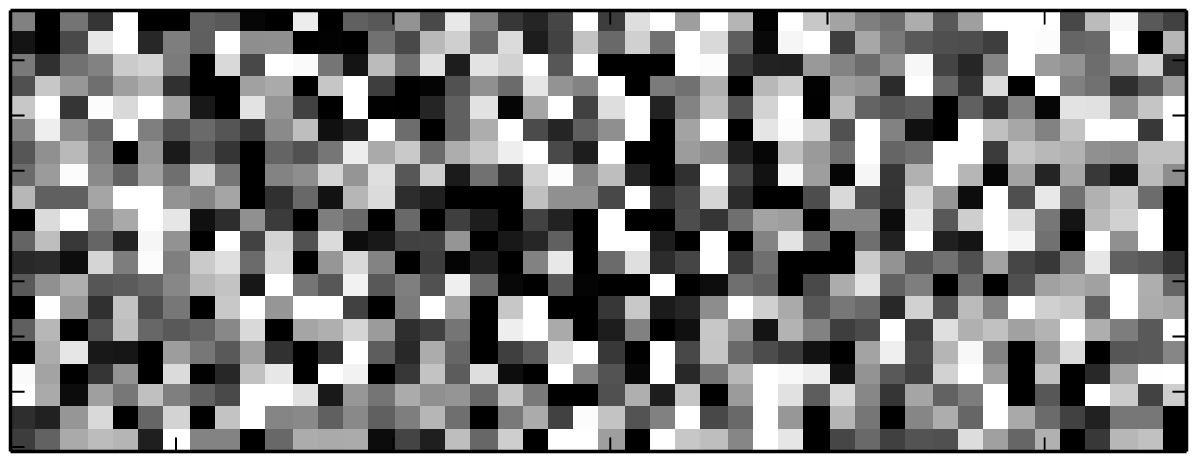}
		\\ 
		\hline
+4.0   & 	\includegraphics[width=0.36\columnwidth]{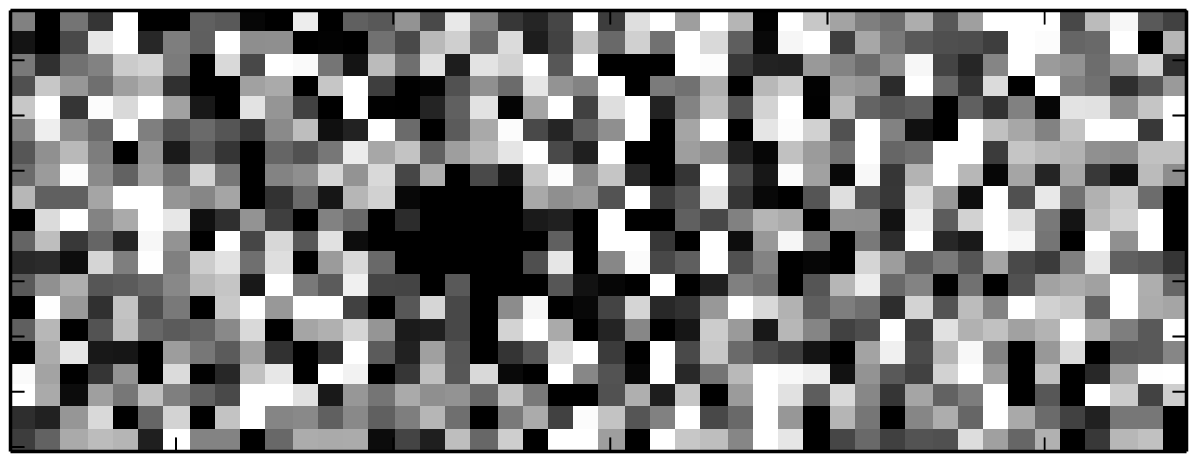} &
		\includegraphics[width=0.36\columnwidth]{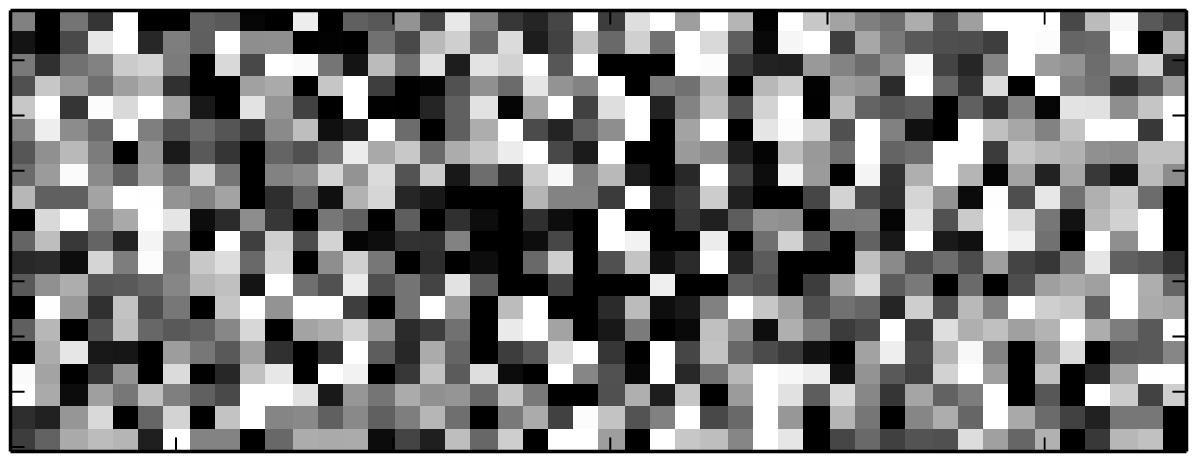}
		\\ 
				\hline
+8.0     & \includegraphics[width=0.36\columnwidth]{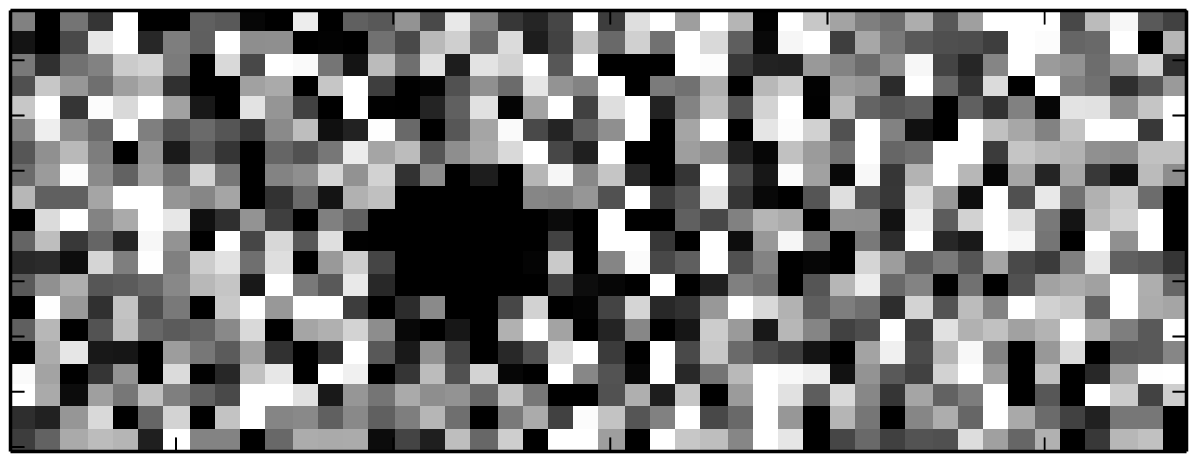} &
		\includegraphics[width=0.36\columnwidth]{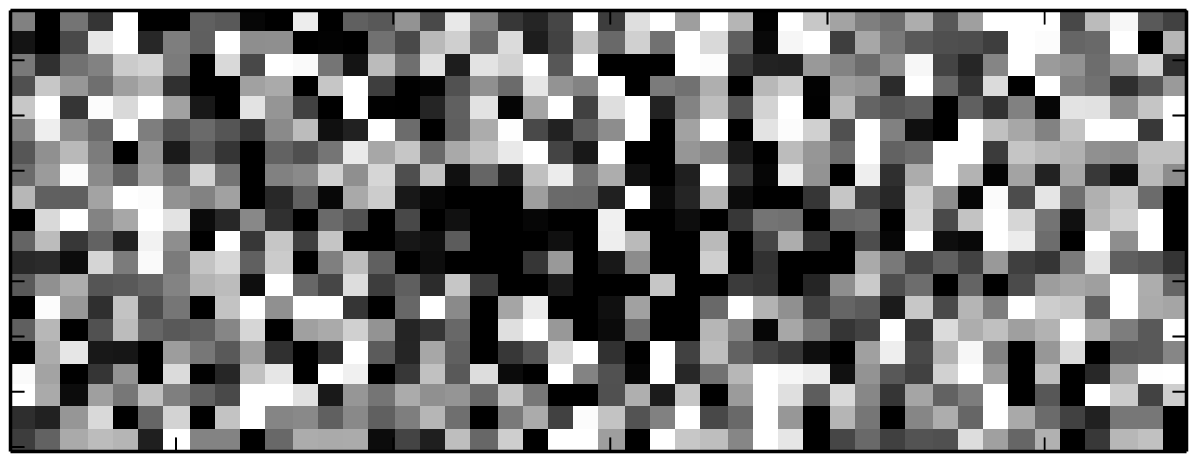}
		\\ 
		\hline

	\end{tabular}

	\begin{tablenotes}
	\item[1] added line flux [$10^{-18}\,\mathrm{erg}\,\mathrm{s}^{-1}\,\mathrm{cm}^{-2}$]

	\end{tablenotes}
	
\end{threeparttable}
	\label{tab:allaboutheii}
	
\end{table}

The production of He\,{\sevensize II}$\,\lambda1640\mathrm{\AA}$ photons, which
originate like Balmer-$\alpha$ in H\,{\sevensize I} from the transition
$n=3\rightarrow2$, requires excitation at least to the $n=3$ level or ionisation with a following recombination cascade. 
Whereas the ionisation of neutral hydrogen, H\,{\sevensize I}, requires only
$13.6\,\mathrm{eV}$, a very high ionisation energy of $54.4\,\mathrm{eV}$ is
necessary to ionise He\,{\sevensize II} and consequently only very 'hard' spectra
can photo-ionise a significant amount. Such hard spectra can be provided by
AGNs, having a power-law SED with significant flux extending to energies beyond
the Lyman-limit, nearly or completely metal-free very young stellar populations with an
initial mass function extending to very high masses, or the radiation emitted
by a shock. 

In Fig. \ref{fig:heii_detlimit}, the 5$\,\sigma$ noise determined as described
in sec.  \ref{sec:detlimits} is shown for the relevant
central-wavelength/box-width space around the expected He\,{\sevensize II}
position and the S/N calculated over the same parameter space is shown in
Fig.~\ref{fig:heii_sn}.  While there might be indication of some excess at
velocities between $-450$ and $0\,\mathrm{km}\,\mathrm{s}^{-1}$, we do not find
sufficient signal to claim a detection.
The two masks giving the highest S/N \footnote{Formally, a very narrow box close
to the strong skylines at $\sim +1000\,\mathrm{km}\,\mathrm{s}^{-1}$ gives a similar high S/N. However, this is clearly affected by residuals from strong
skylines.} are a wider one with a width of $430\,\mathrm{km}\,\mathrm{s}^{-1}$
at a central wavelength of $12453\,\mathrm{\AA}$ ($z=6.591$,
$\Delta v = -160\,\mathrm{km}\,\mathrm{s}^{-1}$) and a narrower one
with a width of $100\,\mathrm{km}\,\mathrm{s}^{-1}$ at a central wavelength of
$12447\,\mathrm{\AA}$ ($z=6.588$, $\Delta v = -300 \,\mathrm{km}\,\mathrm{s}^{-1}$ ).
The S/N in the two cases is after continuum subtraction 1.9 and 1.7,
respectively. 
A somewhat higher S/N can be reached by using a narrower trace more centred on
the position of the expected continuum.  

The 2D spectrum over the relevant velocity offset range is shown in Table
\ref{tab:allaboutheii} in the row labelled 'observed'.
The position of the default trace is indicated as magenta dashed lines in the
smoothed figure, which is identical in the two columns, while the two
extraction boxes giving the highest S/N are indicated by cyan vertical lines in
the two columns, respectively.  Additionally, a sky-spectrum over the same
region and a Gaussian smoothed version are included.  In the Gaussian
smoothing, we excluded pixels being masked in our master high-noise pixel mask.
For a guided eye, it might be possible to identify the excess visually.
Yet, it is certainly possible that noise is seen and it cannot be considered a
detection. It is noteworthy that there is a triplet of weak skylines in the
centre of the region. While these skylines should in principle not increase the
noise by much, as also being consistent with the error spectrum, this would
assume an ideal sky subtraction. As can be seen in the figure, there are however
some unavoidable residuals, extending over the complete slit. 

We tried to understand which line flux would be required for an excess to be
considered visually a safe detections.  We did this by adding Gaussians with
FWHMs of $100\,\mathrm{km}\,\mathrm{s}^{-1}$ and
$400\,\mathrm{km}\,\mathrm{s}^{-1}$ centred on the wavelengths of the two
extraction boxes leading to the highest S/N and assuming a spatial profile as
expected for the continuum. The results are shown in Table
\ref{tab:allaboutheii}.  After visual inspection of four authors, we 
concluded that an additional flux of $2\times10^{-18}$ and
$4\times10^{-18}\,\mathrm{erg}\,\mathrm{s}^{-1}\,\mathrm{cm}^{-2}$ would be
required in the two cases for a detection considered to be safe, corresponding
as expected to approximately $5\sigma$ detections.

Finally, we estimated the $3\sigma$ upper limit on the He\,{\sevensize II}
$EW_0$ in our fiducial $200\,\mathrm{km}\,\mathrm{s}^{-1}$ extraction box,
using the same continuum estimate as used for the continuum subtraction. This
is a continuum flux density at the \ion{He}{II} wavelength of
$4.0\pm0.5\times10^{-20}\,\mathrm{erg}\,\mathrm{s}^{-1}\,\mathrm{cm}^{-2}\,\mathrm{\AA}^{-1}$,
assuming the appropriate slit loss. This results in an upper limit of the
observed frame $EW$ of $75\pm10\,\mathrm{\AA}$, which corresponds to a
rest-frame $EW_0$ of $9.8\pm1.4\,\mathrm{\AA}$. The errors are due to the
uncertainty in $H_{F160W}$, not including the uncertainty in the continuum
slope $\beta$.

\subsection{High ionisation metal lines} \label{sec:hion_lines}

If \emph{Himiko}'s \lya{} emission was powered either by a 'type II' or less
likely by a 'type I' AGN, being disfavoured from the limited
$\mathrm{Ly}\,\alpha$ width, relatively strong C\,{\sevensize IV}$\,\lambda1549$
emission would be expected.  This would be accompanied by somewhat weaker
N\,{\sevensize V}$\,\lambda1240$, C\,{\sevensize III]}$\,\lambda1909$,
He\,{\sevensize II}$\,\lambda1640$,  and Si\,{\sevensize IV}$\,\lambda1400$
emission lines \citep[e.g.][]{VandenBerk:2001cd,Hainline:2011gg}.

\subsubsection{N\,{\sevensize V}} \label{sec:nv}


\begin{figure}
	
	\includegraphics[width = \columnwidth]{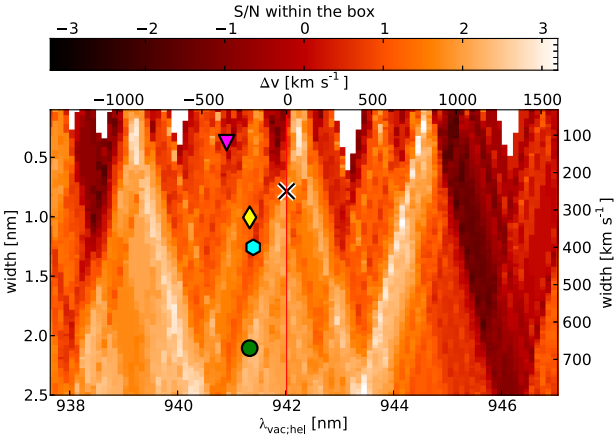}

	\caption{ 
	Similar plot as shown for He\,{\sevensize II} in Fig.
	\ref{fig:heii_sn}.  A telluric correction was applied to the underlying image and a continuum with a flux as motivated in
	sec. \ref{sec:detlimits} was subtracted. The S/N was
	calculated in two boxes at the respective wavelengths of the two
	individual lines ($1238.82\mathrm{\AA}$  and
	$1242.80\mathrm{\AA}$) w.r.t to the blend centre.
			} \label{fig:nv_sn}
\end{figure}

\begin{figure} \centering \begin{tabular}{c} \hline Direct 2d spectrum telluric corrected \\ \hline
	\includegraphics[width = 0.7\columnwidth]{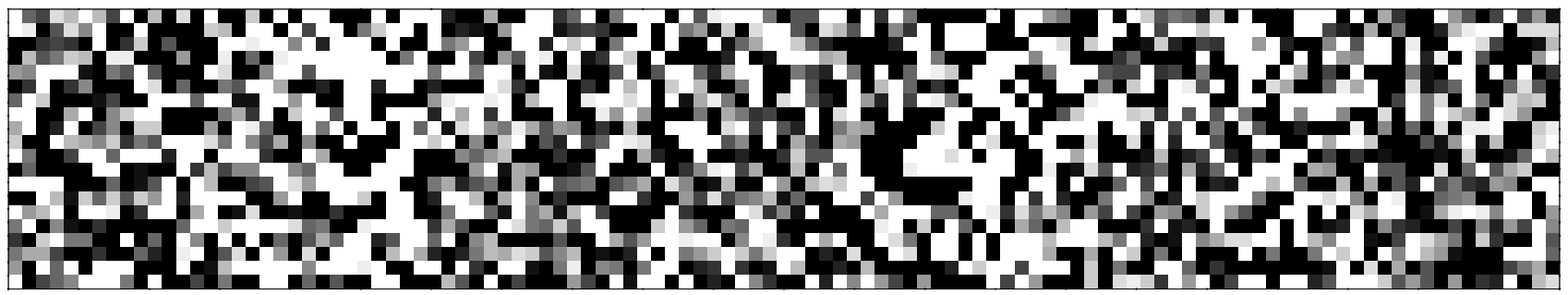}
	\\ \hline Noise map \\ \hline \includegraphics[width =
		0.7\columnwidth]{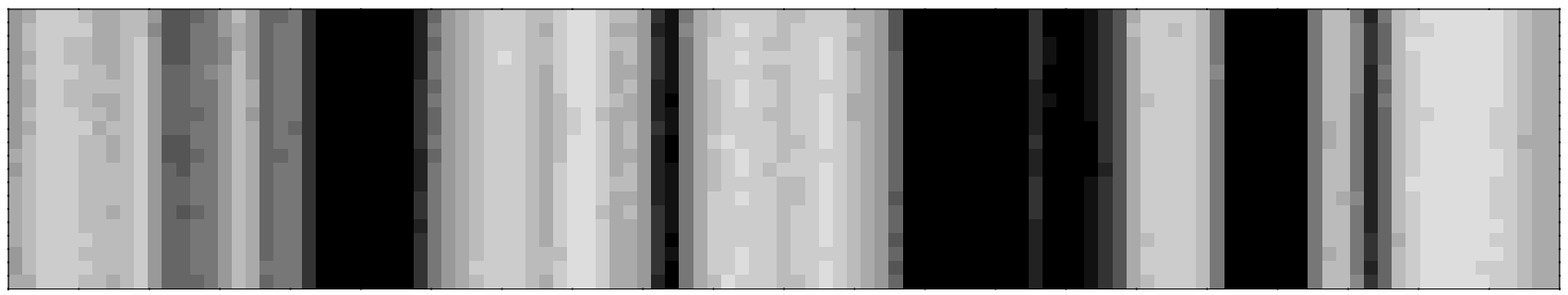} \\ \hline Gaussian convolution of 2d spectrum
		\\ \hline 
			\includegraphics[width =
				0.7\columnwidth]{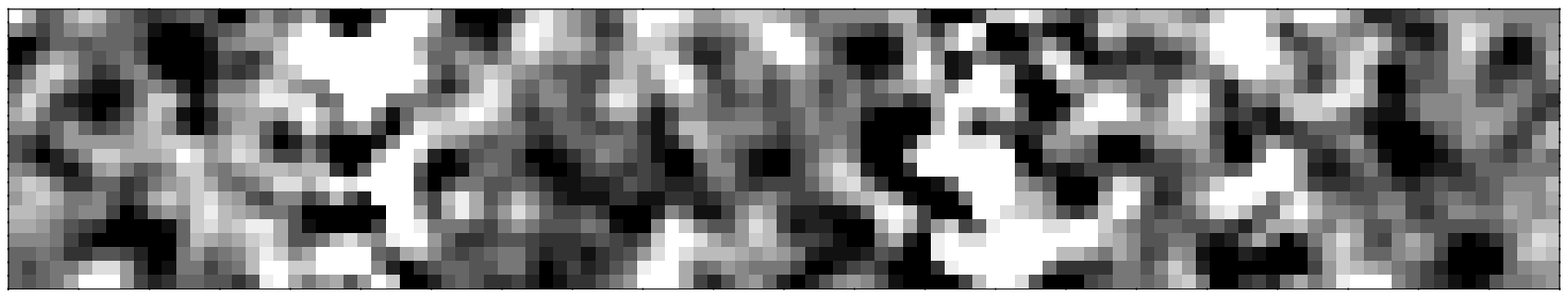}
				\\ \end{tabular} \caption{ Region around the
					N\,{\sevensize V}$\,\lambda 1240$
					doublet, as  expected based on the
					 \lya{} redshift.
					 \textbf{Upper:} Telluric corrected 2D
					 spectrum, scaled between $\pm$ the minimum rms noise, $\sigma_{\mathrm{min}}$, in the shown region. \textbf{Middle:} Noise map linearily scaled between $\sigma_{\mathrm{min}}$  and $3\sigma_{\mathrm{min}}$ (white $<$ black). The high noise regions are both due to skylines and due to regions requiring telluric correction. 
					 \textbf{Lower:} Spectrum smoothed with a Gaussian kernel with a standard deviation of one pixel.}
					\label{smoothnv} \end{figure}


N\,{\sevensize V}$\,\lambda1240$ is a doublet consisting of two lines at
$1238.8\mathrm{\AA}$  and $1242.8\mathrm{\AA}$, respectively.  With their oscillator strength
ratio of $2.0:1.0$, the effective blend wavelength is $1240.2\mathrm{\AA}$.  The
spectrum around N\,{\sevensize V} is shown in the left most panel of Fig.
\ref{fig:multline}.  The hatched wavelength ranges mark the regions for the two
N\,{\sevensize V} lines under the assumption of our fiducial $200\,\mathrm{km}\,\mathrm{s}^{-1}$ wide box.
Weak and blended skylines, which are not marked by our skyline algorithm, and telluric corrected absorption causes the noise to vary strongly over the region.

Using the extraction aperture as shown in Fig. \ref{fig:multline} and subtracting
the continuum in the 2D frame as described in sec. \ref{sec:detlimits}, we
derive for the wider and narrower of our two fiducial boxes excesses of $0.7$ and $0.5\,\sigma$, respectively, where we need to exclude a relatively high fraction of pixels due to high
noise (cf. Table \ref{fig:heii_sn}). The result is consistent with zero.
Exploring the $\Delta v \mbox{--} width$ parameter space for the S/N (Fig.
\ref{fig:nv_sn}), velocity offset and box-width can be chosen in a way
to get a higher S/N.  E.g. boxes at $+450\,\mathrm{km}\,\mathrm{s}^{-1}$  with a width of $840\,\mathrm{km}\,\mathrm{s}^{-1}$ per doublet component, corresponding to a single merged extraction box of $1324\,\mathrm{km}\,\mathrm{s}^{-1}$, have a $S/N$ of 3.4 with an included fraction of 42\%.  However, such a relatively large offset towards the red from the \lya{} redshift seems not feasible.
Restricting the analysis to a more likely range of $\Delta v$
from -500 to $0\,\mathrm{km}\,\mathrm{s}^{-1}$, we would find a maximum
S/N of $2.3$ for box widths of $400\,\mathrm{km}\,\mathrm{s}^{-1}$ at 
no velocity offset w.r.t to \lya{}, with an
included fraction in the box of 45\%.  The flux within non-excluded pixels is
$1.5\times10^{-18}\,\mathrm{erg}\,\mathrm{s}^{-1}\,\mathrm{cm}^{-2}$.
A visual inspection of the relevant region does not allow for the identification of any line (Fig. \ref{smoothnv}).

\subsubsection{Other rest-frame UV emission lines} \label{sec:uvem_lines} 

In Fig. \ref{fig:multline} we are also showing cutouts for two further lines
(\ion{C}{IV} and \ion{C}{III}]).  The relevant wavelength range for the two
components of C\,{\sevensize
IV}\,\,$\lambda\lambda 1548,1551$, which have an oscillator strength ratio of
2.0:1.0, is located in a region of high atmospheric transmittance within the
$J$ band.  While there are a few strong skylines, especially between the two
components, there is enough nearly skyline-free region available. Visually, we
do not see any indication for an excess.
Also statistically, considering again the $\Delta v$ range from $-500$ to
$0\,\mathrm{km}\,\mathrm{s}^{-1}$ w.r.t to the \lya{} peak redshift, we find a
maximum excess of $1.4\,\sigma$ after continuum subtraction.  

By contrast, the C\,{\sevensize III]} doublet is located between the $J$ and
$H$ band, suffering from strong telluric absorption (cf. Fig. \ref{fig:multline}).
Nevertheless, it could be possible to detect some signal in the gaps between absorption. Especially, the relevant part bluewards of the \lya{} redshift
has a relatively high transmission. Both the visual inspection and the formal
analysis of the telluric and continuum corrected spectrum indicate no line.
Detection limits for our fiducial boxes are stated in Table.
\ref{tab:detlimit}.

\ion{Si}{IV}$\,\lambda1403$ is also in a region of high atmospheric
transmission. However, here the overall background noise in the spectrum is
relatively high and as expected for this compared to \ion{C}{IV} usually weak line, we do not
see an excess. 

Another line is N\,{\sevensize IV]}$\,\lambda1486$, which has in rare
cases been found relatively strong both in intermediate and high redshift
galaxies  \citep[e.g.,][]{Christensen2012a, Vanzella:2010jg}.  Unfortunately,
the spectrum is at the expected wavelength for N\,{\sevensize IV]} covered with
strong skylines (not shown).

The same holds for the [O\,{\sevensize III}]$\,\lambda\lambda1661,1666$
doublet, which can be relatively strong in low mass galaxies undergoing a
vigorous burst of star-formation
\citep[e.g.][]{erb2010a,Christensen2012a,Stark2014}.  We do not find any excess
in the relevant wavelength range.  As the region is in addition affected by
several bad pixels, we do not state formal detection limits for this
doublet.

\section{Discussion} \label{sec:discussion}
		
\subsection{SED fitting} \label{sec:sedfitting}

\begin{table}
	\caption{Parameters of the used SSPs, which are the \emph{Yggdrasil} burst models from
		\citet{Zackrisson:2001gs,Zackrisson:2011js} and the BC03
		\citet{2003MNRAS.344.1000B} models combined with nebular emission
		following the recipe of \citet{Ono:2010ed} (\emph{BC03-O10}).} 
	\begin{tabular}{ll}
		\hline
		Parameter & Values \\
		\hline
		\multicolumn{2}{c}{Yggdrasil} \\
		\hline
		\hline
		Metallicities Z ($Z_\odot = 0.02$) & 0.0004, 0.004, 0.008 \\
		&                                    0.02, 0.0 (\ion{Pop}{III}) \\
		IMFs & \citet[$0.1$\mbox{--}$100M_\odot$]{2001MNRAS.322..231K}\\
		& for $Z = 0$: Kroupa, \\
		& log-normal  {\tiny ($1\mbox{--}500\mathrm{M}_{\sun}$,} \\
 &  {\tiny $\sigma = 1\,\mathrm{M}_{\sun}$,  $M_c =
10\,\mathrm{M}_{\sun}$)}, \\
       &  Salpeter ($50\mbox{--}500M_{\sun}$)   \\
		Nebular emission & {\sevensize Cloudy} \citep{Ferland:1998ic}\\
		\multicolumn{2}{l}{Using:
			\citet{Schaerer:2002bm,2005ApJ...621..695V};}\\
			\multicolumn{2}{l}{\citet{Raiter:2010hs}} \\	      
		\hline
		\multicolumn{2}{r}{BC03 (Padova 1994); nebular emission as in \citet{Ono:2010ed}}\\
		\hline
		\hline
		metallicities Z & 0.0001, 0.0004, 0.004, 0.008 \\
		&  0.02, 0.05 \\
		IMF  & \citet[$0.1$\mbox{--}$100M_\odot$]{1955ApJ...121..161S} \\
		Nebular emission & following recipe described \\
		                   & in \citet{Ono:2010ed} \\

		\multicolumn{2}{l}{Using:
			\citet{1980PASP...92..596F,1984ASSL..112.....A,Storey:1995to};}\\
			\multicolumn{2}{l}{\citet{Krueger:1995ty} } \\	                  
		 \hline
		 \hline
		 \multicolumn{2}{c}{Reddening}\\
		 \hline
		 \multicolumn{2}{l}{\citet{Calzetti:2000iy} extinction law,}\\
		 \multicolumn{2}{l}{assuming identical extinction for nebular
		 emission and stellar} \\
		 \multicolumn{2}{l}{  continuum.} \\

	\end{tabular}
	\label{tab:sspinput}
\end{table}

\begin{figure}
	\centering
\includegraphics[width=0.8\columnwidth]{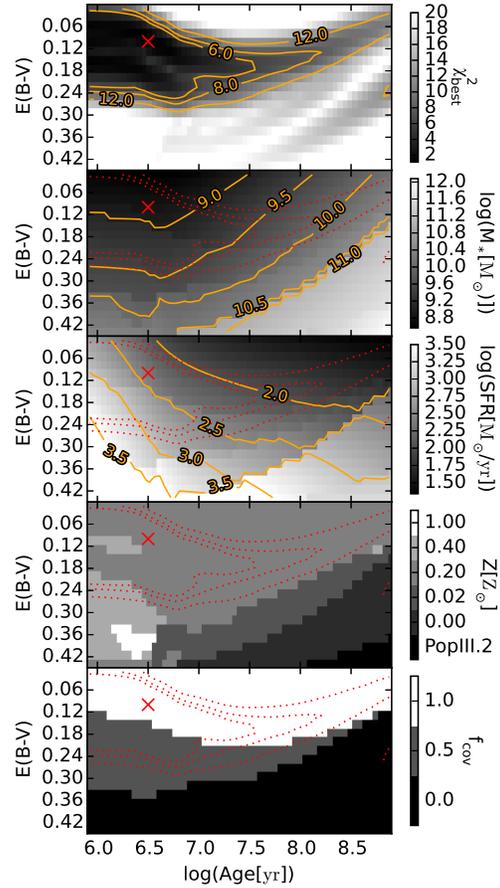}
\caption{\label{fig:fit_yggdrasil_cont} SED fitting results under the assumption
of continuous SFH using the \emph{Yggdrasil} models. Upper panel shows the
lowest possible $\chi^2$ at a given point in the
\emph{age}\mbox{--}\emph{E}(B\mbox{--}V) plain. Four physical quantities for the
best models are shown below. For orientation, the same $\chi_{best}^2$ contours from the top plot are
replicated as dotted line in each of the subplots. The model with the absolute
minimal $\chi^2$ is marked as cross.}
\end{figure}

\begin{figure} \includegraphics[width =
		1.0\columnwidth]{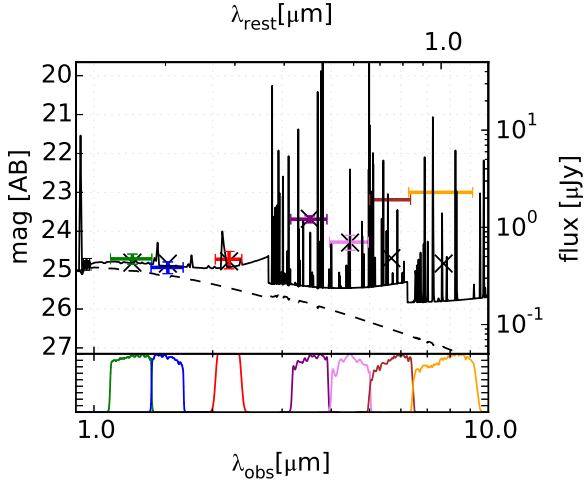}
		\caption{\label{fig:best_sed} Best-fit SED obtained with the Yggdrasil models under
		the assumption of continuous star-formation. For this very young stellar-population ($3\,\mathrm{Myr}$) also the continuum is dominated by the nebular emission, showing the characteristics 'jumps' resulting from the bound-free recombination to different levels in hydrogen. The comparably weak stellar continuum is shown as dashed line. Synthetic magnitudes for the relevant filters are indicated as black crosses. The black circle is the magnitude measured from the spectral continuum.
	}
		 \end{figure} 

We performed SED fitting including $J_{F125W}$, $H_{F160W}$, $K$, and $IRAC
1\mbox{--}4$, in total seven filters, where for $IRAC3\mbox{--}4$ only
upper limits are available. As UKIDSS $K$ and the IRAC data are not resolving
the three components and a profile fit is with the available S/N not feasible,
we fitted the three sources jointly by using the aperture corrected 2\arcsec{}
diameter photometry (Table  \ref{tab:phot_twoarcsec}).  Throughout our SED
fitting we fixed the redshift to $z = 6.590$, a reasonable guess for the systemic redshift based on the \lya{} line.

We used our own {\sevensize python} based SED fitting code {\sevensize
CONIECTO}, which allows both for a MCMC and a grid based analysis. Here, we
derived our results with the grid based option.  The code requires as input
single age stellar populations (SSP) for a set of metallicities.  Additionally,
a pre-calculated nebular spectrum including continuum and line emission needs
to be specified for each age\mbox{--}metallicity pair, and can be added to the
respective SSP with a scale-factor between 0 and 1. This scale-factor can be
understood as covering fraction, $f_{cov}$, or $1-f^\mathrm{esc}_\mathrm{ion}$,
with $f^\mathrm{esc}_\mathrm{ion}$ being the escape fraction of the ionising
continuum. The inclusion of nebular emission has been found crucial for the
fitting of redshift $z\sim6-7$ LAEs and LBGs
\citep[e.g.~][]{Schaerer:2009eq,Schaerer:2010gc,Ono:2010ed}.

SEDs for a given SFH history are obtained by integrating the single age stellar
populations.  We were restricting our analysis to instantaneous bursts and
continuous star-formation. Due to the lack of deep enough rest-frame optical photometry in \emph{IRAC3\mbox{--}4}, which would not be affected by
strong emission lines at our object's redshift, it does not make sense to use
more complicated SFHs.  Already the used set of parameters allows for
over-fitting of the models.

We applied reddening to the integrated SEDs using the \citet{Calzetti:2000iy}
extinction law, assuming the same reddening both for the nebular and the
stellar emission, motivated by evidence for the validity of this assumption in
the high redshift universe \citep{Erb:2006ke}. 

As main input SSPs we used the models 'Yggdrasil' \citep{Zackrisson:2011js},
which include metallcities all from zero (\ion{Pop}{III}) to solar ($Z_\odot=0.02$).
Their code consistently treats both nebular line and continuum emission
\citep{Zackrisson:2001gs} by using {\sevensize Cloudy} \citep{Ferland:1998ic}
on top of stellar populations. We are here referring to SSPs as their publicly
available instantaneous burst models. More details are summarized in Table
\ref{tab:sspinput}.

For reasons of comparison to the SED fitting done by
\citet{Ouchi:2009dp,2013ApJ...778..102O}, we also obtained results with input
models as used in these works. These are based on BC03
\citep{2003MNRAS.344.1000B} models, where a nebular
emission is calculated by the prescription presented in
\citet{Ono:2010ed}.  We are referring in the following to these models as
\emph{BC03\mbox{--}O10}.

After calculating synthetic magnitudes on a large
$age$\mbox{--}$E$(B\mbox{--}V)\mbox{--}$Z$\mbox{--}$f_{cov}$ grid, we have
minimized $\chi^2$ w.r.t to mass for each parameter set. The grid covered ages between
$0\mbox{--}\,800\;\mathrm{Myr}$, with the upper limit being the age of the universe
at $z = 6.59$, and $E$(B\mbox{--}V) from 0.00 to 0.45. Used metallicities were those
available in the input models (Table \ref{tab:sspinput}) and for $f_{cov}$ we
allowed for three different values (0,0.5,1.0).

In Fig. \ref{fig:fit_yggdrasil_cont}, the results are shown for continuous
star-formation using the \emph{Yggdrasil} models.
For each point in the $age$\mbox{--}$E$(B\mbox{--}V) space the $Z$\mbox{--}$f_{cov}$ tuple allowing for
the minimal $\chi^2$, $\chi_{best}^2$, is chosen. The resulting $\chi_\mathrm{best}^2$
contours are indicated in all five subplots, with the five subplots showing $\chi_{best}^2$, stellar mass, star-formation rate averaged over
$100\;\mathrm{Myr}$ ($SFR_{100}$)\footnote{If the age of the stellar population is smaller than $100\;\mathrm{Myr}$, it is averaged over the population age. $SFR_{100}$ is identical to the instantaneous SFR for a constant SFH.}, metallicity, and $f_{cov}$,
respectively. Stellar masses refer to the mass in stars at the point of observation.

The global best fit SED over the explored parameter space is a very young
$3_{-2}^{+32}\;\mathrm{Myr}$ stellar population, which is strongly star-forming ($SFR_{100}$ =
$2.3_{-1.2}^{+26.2}\times10^2\;M_\odot\;\mathrm{yr}^{-1}$) and has a stellar mass of
 $7_{-3}^{+32}\times\;10^8M_\odot$ with a $\chi^2=1.2$.\footnote{Not
 reduced $\chi^2$}
 This best fit model is shown as cross in the maps and as SED in Fig.
 \ref{fig:best_sed}.  The uncertainties on the estimated parameters were
 determined based on the range of models having $\chi^2<6$. A value of six can
 be understood as a good fit for seven filters. 
 
Clearly, a large range of parameters is allowed.  It is only for metallicity
and reddening that relatively strong constraints can be inferred, with mainly
$Z=0.2Z_\odot$ models giving good fits and some $Z=0.4Z_\odot$ models being
allowed within $\chi^2<6$. Models with E$(B\mbox{--}V)$ greater than 0.25 are
unlikely. Further, at least partial nebular contribution is required. 
The wide range of solutions can be understood through the interplay of a
somewhat evolved population with a 4000$\mathrm{\AA}$ break and strong nebular
emission, combined with small amounts of dust extinction. 

The strongest lines in \emph{IRAC1} and \emph{IRAC2} are
[\ion{O}{III}]$\,\lambda\lambda4959,5007$ and H$\alpha$, respectively. The
[\ion{O}{III}] $EW_0$ is expected to peak at metallicities around $Z =
0.2Z_\odot$ \citep[e.g.][]{Finkelstein:2013uf}, strong enough to explain the
blue \emph{IRAC1} - \emph{IRAC2} colour of $-0.6\pm0.2$.

\begin{figure*}
\includegraphics[width=\textwidth]{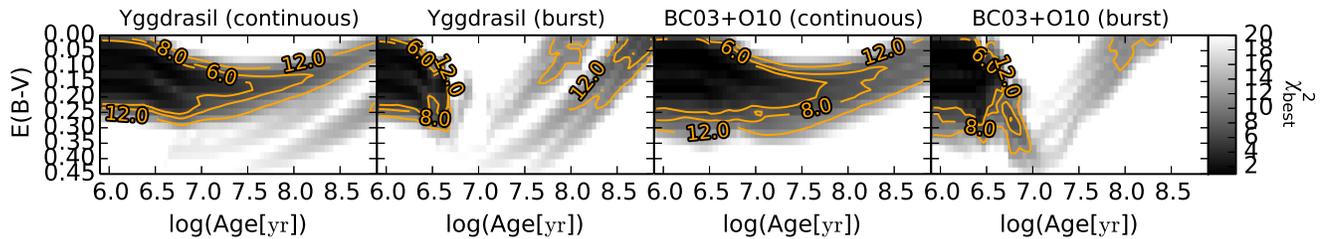}
\caption{\label{fig:fit_sed_diff} The left panel shows the same as the upper
panel in Fig. \ref{fig:fit_yggdrasil_cont}. For comparison, the $\chi_{best}^2$ result
using the \emph{BC03+O10} models and the case of burst SFH is shown.}
\end{figure*}

The $\chi_{best}^2$ plots are compared for burst and continuous star-formation
both using the \emph{Yggdrasil} and the \emph{BC03\mbox{--}O10} model sets in
Fig. \ref{fig:fit_sed_diff}. Burst models give acceptable fits only for very
young ages, where they are basically identical to the continuous star-formation
models.  Both model sets give consistent results, with small differences mainly
existing where specific metallicities are only available in one of the two
sets. E.g., the area of relatively low $\chi_{best}^2$ at high ages for burst
models using \emph{Yggdrasil} is absent in the \emph{BC03-O10} models, as they
require zero metallicity (Kroupa and log-normal IMF). However, these SEDs would
require stellar masses in excess of $10^{11}\,\mathrm{M}_\odot$ and can hence
be considered infeasible. In general, biases are possible due to discretised
metallicities.

We also tested the impact of substituting our $J_{F125W}$ measurement with that
of \citet{2013ApJ...778..102O}.  The fainter $J_{F125W}$ flux shifts the
$\chi^2$ best contours slightly towards higher reddening and hence higher
required SFRs, or alternatively higher ages.  \citet{2013ApJ...778..102O} have
found from their SED fitting with \emph{BCO3\mbox{--}O10} models under the assumption
of continuous SFR a best fit model with a mass of
$1.5_{-0.2}^{+0.8}\times10^{9}M_\odot$, an age of
$4.2^{+0.8}_{-0.2}\times10^6\;\mathrm{yr}$, and an
$E(B-V)=0.15^{+0.04}_{-0.03}$, consistent with our results.\footnote{In the original publication a different best fit model was stated $1.5\times10^{10}M_\odot$, an age of $1.8\times10^8\;\mathrm{yr}$,
$E(B-V)=0.15$, and a SFR of $100\;M_\odot\;\mathrm{yr}^{-1}$; erratum in preparation}

If \emph{Himiko}'s extended \lya{} emission was driven by merger activity between
the three continuum bright objects, SFRs as high as a few
$100\;M_\odot\;\mathrm{yr}^{-1}$ are expected, as seen e.g. in the simulations
by \citet{2013ApJ...773..151Y} focusing on LABs, and consistent with our
best fit model.

Importantly, a burst with the same properties as our best fit model having
happened 100\,Myr before the observed burst, would have faded in all detected
filters by at least a factor 20. Hence, there could easily be a moderately
young stellar population of a few $10^9M_\odot$ without significantly
contributing to the SED.

While the present work was under revision, \citet{Schaerer2015} also published results from SED fitting for \emph{Himiko}. 
Most importantly, they made use of the upper limit on the far-infrared luminosity, which can be estimated from the ALMA upper limit for the $1.2\mathrm{mm}$ (observed-frame) continuum presented by \citet{2013ApJ...778..102O}, in order to constrain the maximum allowed dust extinction. 

Their obtained upper limit corresponds assuming a \citet{Calzetti:2000iy} extinction law to $E(B-V)<0.05\pm0.03$, with the uncertainty due to the unknown dust temperature assumed for the conversion from continuum flux density to FIR luminosity. Using this upper limit, a huge part of the allowed parameter space can be ruled out. 
\citet{Schaerer2015} could argue, using for the SED fitting the photometry of \citet{2013ApJ...778..102O}, that a young and heavily star-forming solution is disvafoured. 
However, due to the somewhat bluer $F125W\mbox{--}F160W$ measured from the CANDELS data compared to that measured in the data of \citet{2013ApJ...778..102O}, our fit requires somewhat less extinction and hence very young models consistent with their upper limit can be found.

\subsection{Further SED considerations}
\label{sec:furthersed}

As shown in section \ref{sec:sedfitting}, the observed broadband magnitudes are
compatible with models over a wide age, mass, and reddening range, where the
single stellar population fit is likely an oversimplification of the problem.
When looking at the $J_{F125W}$ -$H_{F160W}$ colour of the individual
components (cf. Table \ref{tab:phot_comp}), there exists a significant difference
between the two more eastern and the western-most component, with the latter
being about $0.3\;\mathrm{mag}$ redder. This can all indicate differing
stellar populations, differing escape fractions of the ionising continuum,
$f^\mathrm{esc}_\mathrm{ion}$,  or differing amounts of dust.  

Interestingly, the $J_{F125W}$ - $H_{F160W} \sim-0.3\pm0.1$ of the two eastern
components, corresponding to a $\beta = -3.3\pm 0.6$, is somewhat difficult to explain
with a \ion{Pop}{II} when taking account for nebular emission under
the assumption of low ionising continuum escape $f^\mathrm{ion}_\mathrm{esc}$ \citep[e.g.~][]{Bouwens:2010eu}. 
Only for complete escape of the ionising radiation, a population with a metallicity as in our best fit SED ($Z=0.2Z_\odot$) would produce a slope as steep as $\beta=-3$,
whereas for  $f^\mathrm{ion}_\mathrm{esc}=0$ the steepest $\beta$ is a about -2.5.
By contrast, the best fit SED model requires a very low $f^\mathrm{ion}_\mathrm{esc}$ to explain the $IRAC$ magnitudes. 

While each of the components does not differ more than $1.5\sigma$ from this $\beta=-2.5$, the fact that \citet{2013ApJ...778..102O} find at least for the two outer components similar results based on their independent data, increases the probability that the steep slope of $Himiko\mbox{--}E$ is real. 
Certainly, possibilities exist to reconcile the blue colour of the individual components with the low escape fraction required by the SED model. E.g., there could be anisotropic ionising escape in our direction, or the ionising radiation could escape the star-forming regions and the nebular continuum emission is produced somewhat further out, making the nebular emission more extended than the stellar continuum.

Both the visual inspection of Fig.~\ref{fig:cutout} and the spatial profiles
in the individuals OBs (Fig.~\ref{fig:spatial_lyalpha_profiles}) indicate an
offset of the overall continuum light w.r.t to the $NB921$ light distribution, which is dominated by the
\lya{} emission.  The \lya{} emission seems least strong around
$Himiko\mbox{--}W$. 
This has been confirmed by the WFC3 / F098M imaging of
\citet{2013ApJ...778..102O}, who find the strongest \lya{} emission originating
from $Himiko\mbox{--}E$, which is with $EW_{0} = 68^{+14}_{-13}$ however not as high that it would put constraints on the escape fraction.

Two-photon, $2\gamma$, continuum \citep{Breit:1940fw}, emitted by
transitions between the 2s and 1s states of the hydrogen atom, is in the case
of nebular emission powered by a central ionising source one among other
mechanism contributing to the continuum. In the case of cooling radiation,
where the hydrogen atoms are collisionally excited, it would be the sole
continuum contribution, though \citep{Dijkstra:2008ht}. Our data point
provided by the spectrophotometry for the rest-frame wavelength range from
$1238$ to $1285\mathrm{\AA}$ could compared to $F125W$ and $F160W$ resemble the
typical $2\gamma$ dip close to \lya{}. Using the frequency-depended
emissivities as stated in Table 1 of \citet{Spitzer:1951cz}, we calculate the
expected $J_{F125W} - H_{F160W}$ colour for the two photon continuum at
$z=6.590$. We find a value of $J_{F125W} - H_{F160W} = -0.08$. 
Therefore, it cannot be solely responsible for the found very blue $J_{F125W}-H_{F160W}$. In addition, we estimate for the peak flux density expected from
cooling radiation based on the predictions of \citet{Dijkstra:2008ht} and the
measured \lya{} flux for \emph{Himiko} a value of $28.3\,\mathrm{mag}$. This is
more than three magnitudes fainter than the measured flux. Therefore,
$2\gamma$ emission from cooling radiation is unlikely to significantly
contribute to the continuum.

Finally, we note that \cite{Ono:2010ed} have favoured in their SED fitting for the
composite of 91 $NB921$ selected LAEs at $z=6.6$ a very young ($\sim
1\,\mathrm{Myr}$) and very low mass ($\sim 10^8\,\mathrm{M}_{\sun}$) model, with
significant nebular contribution ($f_{\mathrm{esc}}^{\mathrm{ion}} = 0.2$).
They have only included those objects, which are not detected individually in
$IRAC\,1$,\footnote{Based on Spitzer/SpUDS \citep{Dunlop:2007ue}, which is less
deep than Spitzer/SEDS}
 therefore excluding objects like \emph{Himiko}.  The $IRAC\,1$ magnitude in
their median stack is $26.6\,\mathrm{mag}$.  For \emph{Himiko}, the total
magnitude after the slightly uncertain aperture correction is $23.69\pm0.09$.
Simplifying assuming that the flux in \emph{Himiko} is equally distributed
between the three sources, each of them would have a magnitude of $24.9$.  This
is only a factor five higher than the median stack.  Therefore, the individual
components are not as extremely different from the typical $z=6.6$ \lya{}
emitters as the joint photometry suggests.


\subsection{Lyman alpha profile}
\label{sec:lyaprofile}

Due to resonant scattering the \lya{} profile can be modified significantly
both in the ISM/CGM and in the intergalactic medium (IGM). A shaping of the
profile within the ISM/CGM is likely, with \lya{} possibly entering the IGM
with a typical double peaked profile observed for lower redshifts LAEs
\citep[e.g.][]{Christensen2012a,Krogager2013} as predicted by theory
\citep[e.g.][]{Harrington1973,Neufeld1990,Verhamme2006,Laursen2009a}. The blue
or red peak is suppressed in the case of outflows or inflows, respectively. 

On the other hand, \lya{} could leave the CGM also with a nearly Gaussian
profile, as seen in several cases for lower redshift LABs
\citep[e.g.][]{Matsuda:2006gy}, and explainable theoretically by fluorescent
\lya{} emission in a fully ionised halo \citep[e.g.][]{Dijkstra2006a}.

At $z\approx6.5$, the IGM is expected to always suppress the blue part of the
$\mathrm{Ly}\,\alpha$ line completely, while an extended damping wing might
also suppress the red part to some extent, as for example found by
\citet{Laursen:2011ft}, in accordance with previous observational
\citep[e.g.][]{Songaila:2004ie} and analytical results
\citep[e.g.][]{Dijkstra:2007dx}. Therefore, even when leaving the CGM as a
Gaussian, the line might be reprocessed to the observed shape through
scattering in the IGM. We tested the feasibility of this scenario for
\emph{Himiko}.

Assuming that the red slope of the profile is the nearly unprocessed Gaussian,
we fit as a first test a line to this part only, similar to the approach used by
\citet{Matsuda:2006gy}.
We always added to the Gaussian a continuum as measured from the spectrum redwards of \lya{}. 
The wavelength interval used for the fit is shown in Fig. \ref{fig:lyalpha} and ranges from 9236.4 to $9250.0\,\mathrm{\AA}$.

The results from the formal fit are a redshift of $6.589_{-0.013}^{+0.003}$, a
line FWHM of $768_{-91}^{+317}\,\mathrm{km}\,\mathrm{s}^{-1}$ corrected for instrumental resolution, and a line flux
of $2.2_{-0.7}^{+8.1}\times
10^{-16}\,\mathrm{erg}\,\mathrm{s}^{-1}\,\mathrm{cm}^{-2}$ for this full
Gaussian. The best fit flux value would correspond to a $\mathrm{Ly}\,\alpha$
luminosity of $1.7\times10^{44}\,\mathrm{erg}\,\mathrm{s}^{-1}$ after slit loss
correction, and is a factor 3.8 larger than the actually measured one. 

The obtained $FWHM$ is not unrealistically high. \citet{Matsuda:2006gy}
have found for LABs at $z=3.1$ $FWHM$s of $\ga
500\,\mathrm{km}\,\mathrm{s}^{-1}$. These high values were also confirmed by
simulations \citep{2013ApJ...773..151Y}. For high redshift radio galaxies
(HzRGs), the determined width would be still at the very lower limit
\citep[e.g.][]{vanOjik:1997wz}.

Then, we tested whether the IGM absorption could indeed produce the observed
\lya{} profile of \emph{Himiko} by fitting to a model where we apply the median
IGM absorption curve of \citet{Laursen:2011ft}\footnote{reionization starting
at $z = 10$} to a Gaussian, assuming that the Gaussian's peak corresponds to
the systemic redshift and convolving the result with the instrumental
resolution (Fig.~\ref{fig:lyalpha}). For the fit of this combined model, we
used a wavelength range including the full profile, ranging from 9220.3 to
$9250.0\,\mathrm{\AA}$. 

Under these assumptions the best fit underlying Gaussian has a redshift of
$6.59052_{-0.00004}^{+0.00006}$, a line FWHM of
$702_{-13}^{+12}\,\mathrm{km}\,\mathrm{s}^{-1}$, and a line flux of
$1.76_{-0.02}^{+0.02}\times
10^{-16}\,\mathrm{erg}\,\mathrm{s}^{-1}\,\mathrm{cm}^{-2}$, where the
uncertainties do not take account for the range of
possible IGM absorption curves.  The result, which is shown in Fig.
\ref{fig:lyalpha} both before and after applying the median IGM absorption
curve of \citet{Laursen:2011ft}, is surprisingly close to the observed profile
 save some small discrepancy at the peak.  This demonstrates
that IGM absorption alone is a valid option for shaping the \lya{} profile of
\emph{Himiko}.

Summing it up, it is clear that from the observed \lya{} shape alone little can
be concluded about the full profile before entering the IGM.  Therefore, the
fraction of \lya{} scattered out of the line of sight by the IGM is highly
uncertain and upper limits on the flux ratios between rest-frame far-UV lines
and those of \lya{}, as required for the discussion in sec.
\ref{sec:upperlimitsimpl}, need to be treated with caution.

The appropriate loss for \lya{} can range from nearly zero, as would be the
case for a single red peak produced by scattering of \lya{} at an expanding
optically thick sphere, to an enormous fraction in the case that we are only seeing a strongly
suppressed red peak resulting from an infalling medium.  Assuming the Gaussian
as used for the model shown in Fig. \ref{fig:lyalpha} and comparing the flux to
the actually measured one, about 38 per cent would pass the IGM.

As a compromise we state in the following upper limits assuming 50 per cent flux
loss in the IGM or alternatively, and more conservatively w.r.t to upper
limits, zero flux loss to the IGM.  
In addition to the IGM absorption, the \lya{} emission might also be reduced by
dust, even so \emph{Himiko} has been constrained to be very dust poor
\citep{Schaerer2015}.

\subsection{Implications from upper limits on rest-frame far-UV lines}
\label{sec:upperlimitsimpl}

As already outlined in the introduction, there is a range of possible
mechanisms proposed to explain the extended emission around LABs.  The most
popular are photo-ionisation either by a starburst, a (hidden) AGN, or
radiative shocks caused by a burst of \ion{Sn}{II}e following the onset of
star-formation. Alternatively, the powering mechanism could be gravitational
cooling radiation. Possibly, a contribution from several mechanisms is jointly
powering \emph{Himiko}. While our upper limit measurements are not suited to
identify minor contributions, we can compare them to the expectations for the
different mechanisms assuming these as dominating.

For reference, measurements or upper limits on \ion{N}{V}, \ion{C}{IV},
\ion{He}{II}, and \hion{C}{III} from the literature for a selection of
composite spectra and interesting individual LABs and LAEs are listed in Table
\ref{tab:lineratios} and compared to the upper limits obtained by us for
\emph{Himiko}. The stated values are normalised by the respective \lya{}
fluxes. 

As \lya{} and the other lines might have different spatial extent, we
calculated the ratio between these lines and \lya{} for \emph{Himiko} in two
different ways. In both cases we converted before taking the ratio both \lya{}
and the upper limits to aperture corrected values, but while assuming in the
first case that potential emission is as extended as \lya{}, we assumed in the
second case that the emission from other lines is co-aligned with the
continuum. The values for the first case are identical to those one would
obtain from taking the ratios with \lya{} measured in the same apertures\footnote{Meaning same
slit-width and extraction aperture height.} as used for the determination
of the line flux upper limits.

\subsubsection{Stellar population} \label{sec:stell_pop} \begin{figure}
	\includegraphics[width =
		\columnwidth]{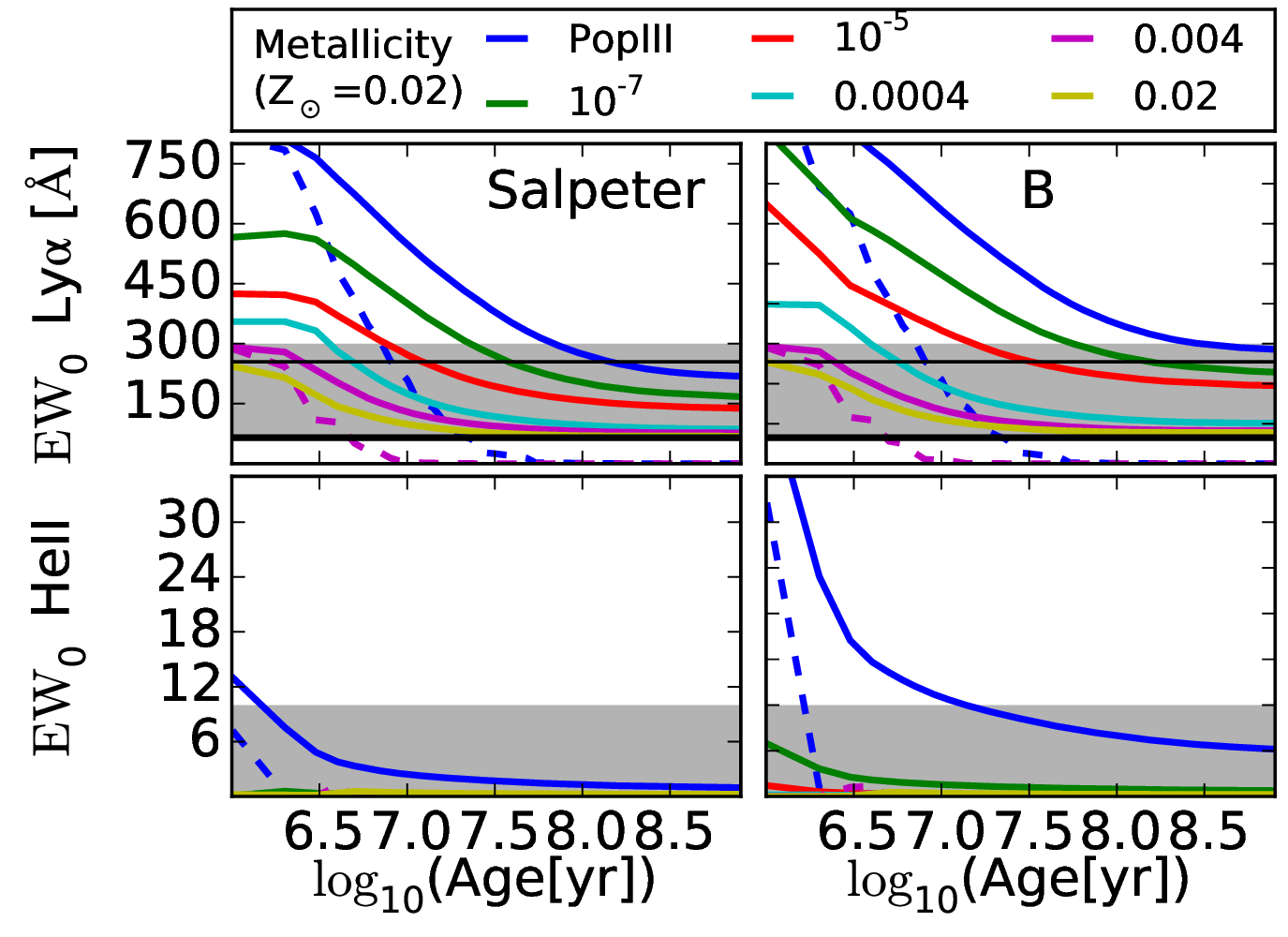}
		\caption{Shown are both the \lya{} and the He\,{\sevensize II}
		$\mathrm{EW}_0$ for two different IMFs assuming constant
		star-formation, where \emph{B} has also a Salpeter slope, but
		includes stars up to $500M_\odot$. We are showing the values
		for 6 different metallicities, ranging from \ion{Pop}{III} ($Z
		= 0$) to solar metallicity ($Z = 0.02$). In addition, burst
		models are included for \ion{Pop}{III} and $Z = 0.004$ as
		dashed lines. Models have larger $EW_0$ with decreasing
		metallicity. The measured \lya{} $EW_0$ is the thick lower
		horizontal line, while the 3$\sigma$ upper limit on
		\ion{He}{II} is shown as a shaded area.  More details are given
		in sec. \ref{sec:stell_pop}} \label{fig:cont_ew} \end{figure}

The range of SED models which give reasonable good fits allows for $SFR_{100}$
ranging from about $100\,M_\odot\,\mathrm{yr}^{-1}$ to extreme
$2600\,M_\odot\,\mathrm{yr}^{-1}$. The \lya{} fluxes and rest frame equivalent
widths for these models are between $6\times10^{-17}$ and
$37\times10^{-17}\,\mathrm{erg}\,\mathrm{s}^{-1}\,\mathrm{cm}^{-2}$ and between
$72$ and $296\,\mathrm{\AA}$. These values were extracted from the
\emph{Yggdrasil} SEDs by using the same approach as for the observation,
meaning a continuum measurement at a rest-frame wavelength of
$1262\,\mathrm{\AA}$. Further, the same extinction law as for the rest of the galaxy was
assumed for \lya{}. These values mean that \emph{Himiko}'s observed \lya{} flux and
$EW_0$ can easily be accounted for by the strong \ion{Pop}{II} star-formation.
Even a relatively strong IGM correction or destruction of \lya{} by dust in the
ISM would not pose a problem.

Still, an interesting question is whether in this heavily star-forming galaxy
\ion{Pop}{III} star-formation might be ongoing $800\,\mathrm{Myr}$ after the
Big Bang or whether our upper limit for \ion{He}{II} allows to rule it out.
The comparison of a measured He\,{\sevensize II} ${EW}_0$ to theoretical models
can put very strong constraints on the allowed IMF-metallicity-age
parameter space.  Combining the information about \ion{He}{II} with an
intrinsic \lya{} ${EW}_0$ would allow to tighten the constraints even further.
However, \lya{}'s susceptibility to resonance effects weakens its usefulness
for this purpose. He\,{\sevensize II} is not affected by this problem. As we do
not detect \ion{He}{II}, we can only test whether we would expect for
\ion{Pop}{III} star-formation \ion{He}{II} flux above our non detection
limit.\footnote{We are referring here as \ion{Pop}{III} to a zero metallicity
population.} 

Using the ${EW}_0$ predictions for \ion{He}{II} and \lya{} calculated by
\citet{Raiter:2010hs}, and following the approach by \citet{Kashikawa:2012fa}
for the Lyman-$\alpha$ emitter SDF-LEW-1, we applied this tool to
\emph{Himiko}.  In Fig. \ref{fig:cont_ew}, the predictions are shown for two
different initial mass functions (IMFs) and six different metallicities ranging
from zero to solar, and assuming constant star-formation. In addition, both for
zero metallicity and $Z=0.004=0.2Z_\odot$ values are included for burst models.
Both IMFs, \emph{Salpeter} and \emph{B}, are power law IMFs with
\citet{1955ApJ...121..161S} slope of $\alpha=2.35$, however differing in the
mass-range with $1\mbox{--}100 M_{\sun}$ and $1\mbox{--}500 M_{\sun}$,
respectively.

In the figure, both the directly observed \lya{} $EW_0$ and the one obtained for
the Gaussian fit to the red wing (sec.  \ref{sec:lyaprofile}) are shown as
horizontal lines. The range of allowed $EW_0$'s from the SED fitting is
indicated as shaded area.  For \ion{He}{II}, the $3\sigma$ upper limit for our
fiducial $200\,\mathrm{km}\,\mathrm{s}^{-1}$ box is represented as shaded area.

Assuming that the intrinsic \lya{} $EW_0$ is the directly measured one, even a
solar metallicity ($Z_\odot = 0.02$) population with standard Salpeter IMF and
continuous star-formation would be independent of the age above the observed
$EW_0$. The approximate correctness of the red wing fit would allow for
significantly stronger constraints. Then, basically only \ion{Pop}{III} and
very metal poor models up to $Z = 10^{-5}$ would be allowed for continuous ages
larger than about $10\,\mathrm{Myr}$. Populations with higher metallicity would
need to be younger. As indicated by our SED fitting, young ages are certainly
possible, and very low metallicities are even in the case of strong
IGM absorption not necessarily required.

For a Salpeter IMF with $1\mbox{--}100 M_{\sun}$, only an extremely young
($<1\,\mathrm{Myr}$) metal free population could produce a He\,{\sevensize II}
${EW}_0$ above the $3\sigma$ detection limit of $9.8\,\mathrm{\AA}$,
independent of IMF, while for the high mass IMF B \ion{He}{II} flux
would be detectable over a longer period at least for a \ion{Pop}{III}. Here,
it is important to note that detectable \ion{He}{II} would be accompanied by
very high \lya{} $EW_0$.  Summing it up, it is only a very young \ion{Pop}{III} 
with an IMF producing very heavy stars, which can be ruled out based on the
\ion{He}{II} upper limit.

\hion{C}{III}, \fion{O}{III}$\,\lambda\lambda1661,1666$, and at least in some
cases also \ion{C}{IV} emission, created by photo-ionisation in the \ion{H}{II}
regions surrounding young high-mass stars, is now understood to be relatively
common in low mass galaxies with high specific star-formation rates and low
metallicities \citep[e.g.][]{Fosbury2003,erb2010a,Christensen2012a,Stark2014}.
While not as strong as in AGNs, the typically strongest of these lines is
\hion{C}{III}, which has been found with $EW_{0}$ up to $\sim15\mathrm{\AA}$
\citep{Stark2014}.\footnote{In the spectrum of the Lynx arc
	\citep[]{Fosbury2003} higher $EW$s have been measured for these
	rest-frame UV lines. However, these extreme $EW$s are not in agreement
	with self-consistent photo-ionisation models and might be a result of
	differential gravitational lensing between illuminated gas and
	illuminating stars \citep[e.g.][]{villar-martin_nebular_2004}.}
A detection of this line seems to be correlated with high \lya{} EW, with the
correlation probably being a result of both line strengths depending
on metallicity \citep[][]{Stark2014}.

Our 3 $\sigma$ upper limit for \ion{C}{III}] $EW_0$ is due to the lines'
location between J and H band with $EW_{0} < 23\mathrm{\AA}$ for the
$200\,\mathrm{km}\,\mathrm{s}^{-1}$ extraction box larger than the observed
values. Only if the lines were close to unresolved and the velocity offset with
respect to peak \lya{} would allow for relatively high transmittance (cf. Fig.
\ref{fig:multline}), a detection would have been feasible.  Similar, one or
both of the [O\,{\sevensize III}]$\,\lambda\lambda1661,1666$ lines, which can
be almost as strong as \hion{C}{III}, would be only for very specific velocity
offsets w.r.t to \lya{} within the skylines gaps.  Therefore, little can be
concluded from these non-detections about the presence of a substantial
population of young massive stars, which in principle could help to break the
degeneracy in the SED fitting.

\subsubsection{Illumination by (hidden) AGN}
Quasars are well known to be responsible for strong \lya{} emission in so
called extended emission line regions, EELRs, spreading in some cases over
several $100\,\mathrm{kpc}$. When being switched on during the phase of cold
gas accretion, \citet{Haiman:2001bb} predicted this emission to originate from
the illumination of primordial gas.  However, at least at lower redshifts, the
illuminated gas is more likely ejected galaxy material driven out by either SNe
or the jets of the AGN itself \citep[e.g.][]{VillarMartin:2007cd}, resulting in
the illumination of metal enriched gas and hence a multitude of emission lines.

With \emph{Himiko's} three spatially distinct components all located within
about $6\,\mathrm{kpc}$ transverse distance and no strong evidence for much
difference in the line of sight direction, they are likely in the process of
merging and hence triggered AGN activity is not unlikely. 
Further, \lya{} emission powered by AGNs is a common and
expected phenomenon.  When considering AGNs, it can be useful both to subdivide
between radio-quiet and radio-loud \citep[e.g.][]{McCarthy:1993ek}, and between
obscured ('type II') and non-obscured ('type I') by a dusty torus.  Extended
emission has been found for all cases \citep[e.g.][]{Matsuoka:2012dn}. In the
following we are discussing all four options for \emph{Himiko}, especially
w.r.t to our non-detection limits.

As discussed by \citet{Ouchi:2009dp}, neither of the existing X-ray
(XMM-Newton), $24\mu m$ (\emph{Spitzer} MIPS), and sub-mm (850$\mu m$ SCUBA)
data is deep enough to rule out a type I quasar, when scaling the
\citet{Elvis:1994du} quasar template to the rest-frame VIS $IRAC 1$ flux.
However, there is a strong argument against a type I AGN in the \lya{} width,
which is even after accounting for IGM absorption too narrow to be originating
from a type I AGN (cf. sec. \ref{sec:lyaprofile}).  Another argument against a
quasar, at least for the two eastern components, is the continuum slope.
\cite{Davis:2007bu} find for SDSS quasars a mean slope of $\beta =
-1.63$\footnote{Converted from the $\nu^\alpha$ as stated in
	\citet{Davis:2007bu} to $\lambda^\beta$}
and very few objects with slopes bluer than $\beta = -2.0$ for the wave-length
range between $1450\mathrm{\AA}$ and $2200\mathrm{\AA}$.  Although the
redshifts of this sample is limited to $1.67 \le z \le 2.09$, we do not
expected the quasar spectra to be much bluer for higher redshifts.  Therefore, it is unlikely that the two eastern components are dominated by direct continuum emission
from the accretion disk.  Finally, our $3\,\sigma$ upper limit on
\ion{C}{IV}/\lya{} of 0.13, using the $600\,\mathrm{km}\,\mathrm{s}^{-1}$
aperture and assuming a crude IGM correction of a factor two is two times below
the typical value for type I quasars in the composite spectrum of
\citet{VandenBerk:2001cd}.  Certainly, this comparison falls somewhat short, as
a \ion{C}{IV} width of $600\,\mathrm{km}\,\mathrm{s}^{-1}$ is too low for a
type I quasar. Much higher widths are not justifiable based on our
observed \lya{} line width.

By contrast, \emph{Himiko's} \lya{} width is an option for the extended
emission line regions around quasars, may it be either a powerful high redshift
radio galaxy (HzRG) or a radio quiet type II quasar. While typical \lya{} line
widths of $\sim1000\,\mathrm{km}\,\mathrm{s}^{-1}$ seem somewhat high compared
to \emph{Himiko}'s measured one, it might be possible to explain such a width
under the assumption of strong IGM absorption (cf. sec.  \ref{sec:lyaprofile}).
Further, it is known that HzRGs with very relaxed kinematics exist. MRC
0140\mbox{--}257 has over the complete nebula \lya{} $FWHM \la
500\,\mathrm{km}\,\mathrm{s}^{-1}$ \citep{villarmartin2007a}, with very narrow
emission in the peaks of \lya{} surface brightness ($FWHM \sim
250\,\mathrm{km}\,\mathrm{s}^{-1}$).  Also, while the extended \lya{} emission
around HzRGs is usually spatially kinematically disturbed by several $100
\,\mathrm{km}\,\mathrm{s}^{-1}$, MRC 0140\mbox{--}257 has comparable low
velocity offsets. \emph{Himiko} would need to be similar to this rather special
object, as we see in our stacked spectrum no evidence for strong velocity
gradients (cf. both Fig. \ref{fig:nb_zprime} and \ref{fig:all_lyalphapos}).
We remark that velocity gradients could be smoothed out in our stack, as it is a superposition of observations with different slit angles.
However, also in the spectrum of \citet{Ouchi:2009dp}, who observed with a
fixed slit position, only a small \lya{} velocity gradient of
$60\,\mathrm{km}\,\mathrm{s}^{-1}$ is found over the east-west direction.

Comparing the rest-frame UV and optical photometry, we find that the magnitudes
measured for \emph{Himiko} are indeed values reasonable for those expected from
HzRGs, when taking the strong and extended emission in the EELRs into account.
Assuming a flat continuum in $f_\nu$ and $H_{\mathrm{F125W}}$
($24.93\,\mathrm{mag}$) as normalisation, we can estimate the flux due to
emission lines in IRAC1 and IRAC2 to
$14.8\times10^{-17}\,\mathrm{erg}\,\mathrm{s}^{-1}\,\mathrm{cm}^2$ and
$4.8\times10^{-17}\,\mathrm{erg}\,\mathrm{s}^{-1}\,\mathrm{cm}^{-2}$,
respectively, corresponding to a ratio of 3.1. Summing up the relevant line
fluxes in the HzRG composite spectrum of \citet{Humphrey:2008jv}, a typical
emission line flux ratio of 3.0 between IRAC1 and IRAC2 would be obtained,
\footnote{The relevant lines in IRAC1 are \ion{He}{II}$\,\lambda 4686$,
H$\mathrm{\beta}$, and [\ion{O}{III}]$\,\lambda\lambda 4959,5007$ and in IRAC2
\ion{O}{I}$\,\lambda 6300$, H$\mathrm{\alpha}$, and
[\ion{N}{II}]$\,\lambda\lambda6548,6583$.} consistent with that measured for
\emph{Himiko}.  Assuming that about 70 per cent of the line flux in IRAC1 is
due to [\ion{O}{III}]$\,\lambda5007$, the [\ion{O}{III}]$\,\lambda5007$
rest-frame $EW_0$ would be 1400$\,\mathrm{\AA}$. Such a value is not
unrealistic for HzRGs \citep[e.g.][]{Iwamuro2003a}.  Finally, based on the
[OIII] flux estimated above, \emph{Himiko}'s \lya{} flux would be using the
\citet{Humphrey:2008jv} average ratio expected to be a factor 1.7 higher than
the actually observed one.  This factor is easily within the uncertainty of the
IGM absorption correction.

While this similarity in line ratios is interesting, it is not entirely
surprising when considering our best fit stellar SED model.
\citet{Humphrey:2008jv} conclude, that the line emission in these HzRGs is not
mainly powered by the interaction of the radio lobes with the gas, but rather
by photo-ionisation, and that the illuminated gas has typical metallicities of
$Z=0.2Z_\odot$.

A means to discern between an AGN or a young stellar population as ionising
source would be through the strength and ratios of rest-frame far-UV
emission lines.  The ratios between these lines and \lya{} are in general
expected to be stronger, when being powered by an AGN.

A complicating factor
when interpreting non-detection limits is the extent of the possible line
emission. Extended \lya{} emission could be a consequence of resonant
scattering or of in situ production by ionising radiation escaping the ISM.  In
the first case other emission lines would be almost co-aligned with the UV
continuum, while in the latter case they could be spatially extended as the
\lya{} emission \citep[e.g.][]{Prescott2015a}, resulting for our object in a
larger slit loss.

The typical ratio between \ion{C}{IV} and \lya{} in the \citet{Humphrey:2008jv}
composite is with 0.15 similar to the 3$\sigma$ upper limit for our target,
when assuming the $600\,\mathrm{km}\,\mathrm{s}^{-1}$ extraction box and a
crude \lya{} IGM correction of a factor two, which corresponds to
\ion{C}{IV}/\lya{} $< 0.13$ and $<0.18$ assuming a spatial extent similar to
the continuum or the \lya{} emission, respectively.  Also, the typical
\ion{He}{II}/\lya{} of $0.09$ would be detected with at least $3\sigma$ even
for \ion{He}{II} emission as extended as \lya{}, assuming again the crude IGM
correction and the wider extraction box. The advantage of \ion{He}{II} compared to \ion{C}{IV} is that
it would be present even when assuming primordial composition of the
illuminated gas.

Another problem with the interpretation of \emph{Himiko} as a HzRG is the available
$100\,\mu\mathrm{Jy}$ 1.4 GHz VLA upper limit \citep{Simpson2006a}.  Assuming
example radio spectral indices of $-0.5$, $-1.0$, and $-1.5$, with ultra steep
slopes more typical for high-z galaxies, this corresponds to $L_{1.4GHz}$ of
$1.8\times10^{25}$, $5.0\times10^{25}$,
$1.4\times10^{26}\,\mathrm{W}\,\mathrm{Hz}^{-1}$.  Converted to
$5\;\mathrm{GHz}$ the upper limit would be around
1\mbox{--}2$\times10^{25}\,\mathrm{W}\,\mathrm{Hz}^{-1}$ and hence at the very
lower limit of what would still be considered a radio-loud galaxy according to the classification by
\citet{Miller1990a}.  On the other
hand, if the main ionising source for the EELRs is photo-ionisation by the
quasar and not the radio jets, the non detection in the radio data is not
ruling out the possibility of a hidden AGN as powering source. Indeed, EELRs
have been found around the less known population of radio-quiet type II quasars
\citep[e.g.][]{ghandi2006a, villarmartin2010c}.

For several among the $z \approx 2 \mbox{--} 3$ LABs, a hidden AGN has been
identified at least as a partial contributor to the \lya{} ionising flux. E.g.,
the blob of \citet{Dey:2005dla} at $z\approx 2.7$ shows clear indication of a
dust-enshrouded AGN, producing both He\,{\sevensize II} and C\,{\sevensize IV}
emission with relatively narrow width
($\sim365\;\mathrm{km}\;\mathrm{s}^{-1}$).  Scaling the strength of the lines
to our \lya{} surface brightness, both lines should be detected.

 For the narrow-lined AGNs among those UV selected galaxies by
 \citet{Steidel:2004bd} at $z\sim2\mbox{--}3$ \citep{Hainline:2011gg}, the
 ratios between the rest-frame far-UV lines and \lya{} would be even higher.

In addition to the constraints presented here, \cite{Baek:2013ko} have 
argued for a criterion to discriminate between LABs being powered by either
star-formation, Compton-thin, or Compton-thick AGNs based on the combined
information about the observed surface brightness profile and skewness of the
\lya{} line.  They seem to conclude that \emph{Himiko} is not in the right
region of the parameter space for having an AGN as source.

\subsubsection{Gravitational cooling radiation} \label{sec::cooling_radiation}

Early semi-analytical models assumed that the gas feeding a halo is heated to
the virial temperature. However, cosmological SPH simulations indicate that the
majority of the infalling gas might never reach these high temperatures
\citep{Fardal:2001ds}.  Under the assumption of primordial element composition,
meaning basically only H and He, the major cooling channel of the gas is
consequently not through Bremsstrahlung, as for gas with temperatures above
$10^6K$, but through collisional excitation of hydrogen and, depending on the
temperature, also through He\,{\sevensize II}.  Collisional excitation cooling
peaks at $10^{4.3}\mathrm{K}$ and $10^{5}\mathrm{K}$ for H\,{\sevensize I} and
He\,{\sevensize II}, respectively. 

For several observed LABs in the literature, it has been suspected that cooling
radiation is the major driver of the extended \lya{} emission.  For instance,
\citet{Nilsson:2006hi} had favoured this scenario for their LAB at $z=3.16$, as
they could not identify an obvious counterpart, even in deep GOODS \emph{HST} imaging.
Recent reanalysis of this object based on the extended availability of multi-wavelength data allowed to identify one of the objects in the field, which is located at a distance of about $30\;\mathrm{kpc}$ from peak \lya{} emission, as an obscured AGN \citep[][]{Prescott2015a}.
Unfortunately, no spectrum deep enough to seriously probe He\,{\sevensize II} and the other rest-frame far-UV lines is available for this object.

A LAB, where He\,{\sevensize II} has been observed, is the one by
\citet{Scarlata:2009ce}. While an AGN can also be identified photometrically
within this object, they favour gravitational cooling radiation being at least
partially responsible for He\,{\sevensize II} because of the non-detection of
C\,{\sevensize IV}.

The He\,{\sevensize II} luminosity of
$8\times10^{41}\,\mathrm{erg}\,\mathrm{s}^{-1}$ within their extraction mask
would correspond at $z=6.595$ to a flux of
$1.6\times10^{-18}\,\,\mathrm{erg}\,\mathrm{s}^{-1}\,\mathrm{cm}^{-2}$ , a flux
which would not be safely detectable by us. On the other hand, scaling
\ion{He}{II} to the higher \lya{} flux in \emph{Himiko}, we would
be able to significantly detect it (cf. Table \ref{tab:lineratios}).

The question is whether the corresponding high \ion{He}{II} luminosity would be
at all feasible for cooling radiation and whether a substantial amount of the
\lya{} emission could originate from cooling radiation. Insight can be gained by
comparison to published results from simulations.

As a relatively narrow line is predicted for \ion{He}{II} in the case of cooling radiation \citep{Yang:2006hv}, we compare to the $3\sigma$ upper limit of our standard
$200\,\mathrm{km}\,\mathrm{s}^{-1}$ aperture. This limit corresponds to a
luminosity of
$L_{\mathrm{\ion{He}{II}}}=1.5\times10^{42}\,\mathrm{erg}\,\mathrm{s}^ {-1}$, a
value high compared to the luminosities of even the heaviest halos in the
simulations of \citet{Yang:2006hv}. Here, it needs to be noted that the results of \citet{Yang:2006hv} are for $z=2.3$.

\citet{Yajima:2012wy} have run a hydrodynamical simulation combined
with \lya{} radiative transfer, including all radiative cooling and heating,
star-formation in a multi-phase ISM, stellar feedback, and UV background. Their
simulation traces a halo developing a Milky Way size galaxy over cosmic times.
We have extracted from their presented evolution plots several quantities at
redshift $z = 6.6$. A fraction of 65 per cent of \lya{} is predicted by them to
be due to cooling radiation, mainly created within the cold accretion streams,
where the total $\mathrm{Ly}\,\alpha$ luminosity is
$L_{\mathrm{Ly\alpha}}=4.5\times10^{42} \mathrm{erg}\,\mathrm{s^{-1}}$.
Therefore, even before correcting for IGM absorption and before applying a
surface brightness threshold, the \lya{} luminosity is more than a factor ten
lower than \emph{Himiko}'s \lya{} luminosity.

Taken this and the fact that the SED fitting suggest vigorous star-formation,
it is unlikely that a major fraction of \emph{Himiko's} \lya{} luminosity is
produced by cooling radiation. Therefore, as each particle in the simulations of 
\citet{Yang:2006hv} has $f_{He\,{\sevensize II}\lambda1640} \le 0.1\times
f_{\mathrm{Ly}\alpha}$, the \ion{He}{II} flux is not very likely to be
detectable and no strong conclusions can be drawn from the non-detection.

\subsubsection{Shock ionisation} \label{sec:shock_ionization}

Another possible explanation for strong and extended \lya{} emission are large scale shock-super-bubbles
\citep{Taniguchi:2000ba}, which might be the consequence of galactic winds
powered by a large number of core-collapse supernovae within the first few $100\,\mathrm{Myr}$
after the onset of the initial starburst \citep{Mori:2004kw,Mori:2007il}.  Assuming that the
kinetic energy of the supernova ejecta is converted into radiation by fast-shocks and
combining information on C\,{\sevensize
IV}/He\,{\sevensize II} and N\,{\sevensize V}/He\,{\sevensize II} ratios with a measured line width, one could get insight into the feasibility of this mechanism as contributing power source for the extended \lya{} emission through comparison to shock model grids \citep[e.g.~][]{Allen:2008gj}. In absence of a significant detection of at least one of the lines, such an analysis is not feasible. 

 \begin{table*}
	\caption{
	Line ratios of He\,{\sevensize II}, N\,{\sevensize V}, C\,{\sevensize III}], and C\,{\sevensize IV} to \lya{} as found in samples and
	individual objects by various studies. Upper limits are
	$3\,\sigma$.
	References: \textbf{$[0]:$} Line ratios as measured in this
	study for \emph{Himiko}. 
The ratios are calculated based on slit loss corrected fluxes. While we have for \lya{} assumed a spatial extent given by the narrowband NB921 image, the other lines were corrected under the assumption that they are distributed as the continuum. If they were as extended as the resonant \lya{} line, the upper limits would be a factor 1.5 and 1.4 less stringent for N\,{\sevensize V} and the three lines in the NIR arm, respectively.
 The non-detection limits are stated for our two fiducial boxes. Values in brackets are assuming a crude IGM correction of a factor two. 
 \textbf{$[1]:$} SDSS quasar composite spectrum \citep{VandenBerk:2001cd} \textbf{$[2]:$} Composite
 spectra for AGNs among rest-frame UV selected galaxies \citep{Hainline:2011gg}. The
 sample is split into those with high and low \lya{} $EW_0$. To convert the values 
 and the spread in the equivalent widths stated by \citet{Hainline:2011gg} to values and
 spreads for the fluxes, we have used the continuum slopes of $\beta =
 -0.1\pm0.4$ and $\beta = -0.5\pm0.3$ stated by them for the composite spectra
 of the high and low $EW_0$ sample, respectively. 
 \textbf{$[3]:$} Sample of
 nine high-redshift radio galaxies \citep{Humphrey:2008jv} \textbf{$[4]:$}
 \citet{Prescott:2009ew} Values are for their night 1. \lya{} flux based on same
 aperture as other lines.  \textbf{$[5]:$} \citet{Dey:2005dla}  \lya{} flux is
 based on same aperture as other lines.  \textbf{$[6]:$} \citet{Scarlata:2009ce}
 referring to component \emph{Ly1}; \lya{} flux based on same aperture as other
 lines.  \textbf{$[7]:$} 1$\,\sigma$ He\,{\sevensize II} detection from
 \emph{HST}/WFC F130N NB imaging \citep{Cai:2011hc} \textbf{$[8]:$}
 He\,{\sevensize II} non-detection for extreme \lya{} $EW_0 \approx 900
 \mathrm{\AA}$ object \citep{Kashikawa:2012fa} \textbf{$[9]:$} He\,{\sevensize
 II} non-detection for SDF J132440.6+273607 \citep{Nagao:2005bn} \textbf{$[10]:$} $10^9\,M_\odot$ LAE at z=2.3, showing strong \ion{He}{II} and \ion{C}{III}] emission likely powered by star-formation \citep{erb2010a}.
 }
 \label{tab:lineratios}

 \begin{threeparttable}
	 
\begin{tabular}{lccccc} 
	\hline 
	& \lya{} & N\,{\sevensize V} &  C\,{\sevensize IV} &  He\,{\sevensize II} & C\,{\sevensize III}]  \\
	\hline
	& \multicolumn{5}{c}{\textbf{Himiko}}\\
	\hline
	& \multicolumn{5}{c}{Based on aperture corrected line fluxes and upper limits without (with) IGM correction.}\\
	\hline
	box: $200\,\mathrm{km}\,\mathrm{s}^{-1}$; $v_{\mathrm{off}} = -250\,\mathrm{km}\,\mathrm{s}^{-1}$ & 1.00& $<0.04$ ($<0.02$) & $<0.15$ ($<0.07$) & $<0.07$ ($<0.03$) & $<0.12$ ($<0.06$) \\
	 box: $600\,\mathrm{km}\,\mathrm{s}^{-1}$; $v_{\mathrm{off}} = -250\,\mathrm{km}\,\mathrm{s}^{-1}$ & 1.00&  $<0.08$ ($<0.04$) & $<0.25$ ($<0.13$) & $<0.10$ ($<0.05$) & $<0.18$ ($<0.09$) \\

	 \hline \hline
	 
	 \hline & \multicolumn{5}{c}{\textbf{Composite spectra}}\\
	 \hline 
	 Broad-line AGN $[1]$ &  1.00  & 0.03 &   0.25
	& 0.005 & 0.16     \\
	 Narrow-line AGN $[2]$  & $1.00$ &
	 $0.05\pm{0.01}$ &  $0.20\pm{0.04}$ & $0.07\pm{0.02}$ & $0.16\pm{0.07}$  \\
	 \hspace{2ex}
	 {$EW_{Ly\alpha} > 63\,\mathrm{\AA}$} &&& \\
	 Narrow-line AGN
	 $[2]$ & $1.00$ & $0.16\pm{0.05}$ &  $0.25\pm{0.09}$ & $0.17\pm{0.06}$ & $0.17\pm{0.09}$ \\
	 \hspace{2ex}{$EW_{Ly\alpha} < 63\,\mathrm{\AA}$} &
	 &  \\
	 Radio-galaxies ($z\sim2.5$) $[3]$ & 1.00 & 0.04 & 0.15  & 0.09 & 0.06 \\
	 \hline
	 & \multicolumn{4}{c}{\textbf{Example LABs}} \\
	 \hline
	 LAB PRG1 ($z=1.67$) $[4]$ & 1.00 &  $<0.39$ &  $<0.09$
	 & 0.12 & $<0.07$ \\
	 LAB ($z=2.7$)	 $[5]$    & 1.00 
	 & $<0.02$  &     0.13  & 0.13 & 0.02  \\
	 LAB ($z=2.38$) $[6]$ & 1.00  & N/A & $<0.45$  & 0.36 & N/A \\
	 \hline
	 & \multicolumn{5}{c}{\textbf{Example LAEs}} \\ 
	 \hline 
	 LAE IOK ($z=6.96$) $[7]$ N/A & $1.00$ & N/A & N/A  & $0.06\pm{0.05}$ & N/A
	\\ 
	 LAE SDF-LEW-1 ($z=6.5$) $[8]$  &  1.00 & N/A   &    $<0.013$ & 
	 $<0.015$ & N/A  \\
	 LAE SDF ($z=6.33$) $[9]$ &  1.00 & N/A   &  $<0.675$ & 
	 $<0.340$ &  N/A\\
	 BX418 $[10]$ & 1.00 & N/A & abs & 0.034 & 0.048 \\
 \end{tabular}

\end{threeparttable}

\end{table*}

\section{Conclusion}

With our analysis we add further hints to the puzzle of what is powering the
remarkable \lya{} emitter \emph{Himiko}.

First, we have detected a continuum in the spectrum showing a clear break at
the wavelength of \lya{}, ruling out any remaining chance of it being a lower
redshift interloper. The fact that the continuum appears like a single step
function indicates that there are no large velocity differences between the
three distinct UV bright components.

From SED fitting, including CANDELS $J_{F125W}$ and $H_{F160W}$ data, we argue
for a young and heavily star-forming stellar population, with a total stellar
mass of the order $10^9M_\odot$ and a metallicity of $Z=0.2Z_\odot$, with the
bright \emph{IRAC1} magnitude explained by very strong [\ion{O}{III}] emission.
While we find that similar broadband magnitudes would also be produced by lines
in the extended emission line regions around high redshift radio galaxies, this
scenario is for several reasons more unlikely.  Among them is the most
important result of this work.  Our upper limits on important rest-frame far-UV
lines are clearly disfavouring an AGN as sole powering source for the extended
\lya{} emission. However, due to the natural lack of knowledge about the appropriate
slit loss, line-width, and systemic redshift and the spread in line strength for
AGNs, it is not entirely impossible that an AGN has escaped our
detection.

\section*{Acknowledgments} Based on observations made with ESO telescopes at the
La Silla Paranal Observatory under programme ID 087.A-0178(A).  The Dark
Cosmology Centre is funded by the DNRF.  JPUF and PL acknowledge support from
the ERC-StG grant EGGS-278202.  This work is based on observations taken by the
CANDELS Multi-Cycle Treasury Program with the NASA/ESA \emph{HST}, which is
operated by the Association of Universities for Research in Astronomy, Inc.,
under NASA contract NAS5-26555. 
This research made use of Astropy, a community-developed core Python package
for Astronomy (Astropy Collaboration, 2013). We thank Martin Sparre for
providing us with his Python XSHOOTER pipeline manager, and Nobunari Kashikawa and Yoshiaki Ono for providing us with information.

\label{lastpage}

\bibliographystyle{mn2e} \bibliography{library_papers_m}

\end{document}